\documentclass[twocolum]{pasj01}

\begin{document}
\SetRunningHead{et al.}{Dust Emission from the CS dust in SC SN 2012dn}

\title{OISTER Optical and Near-Infrared Observations of the Super-Chandrasekhar Supernova Candidate SN 2012dn: Dust Emission from the Circumstellar Shell}

\author{Masayuki \textsc{Yamanaka}\altaffilmark{1,2},
        Keiichi \textsc{Maeda}\altaffilmark{3,4},
        Masaomi \textsc{Tanaka}\altaffilmark{5}, 
        Nozomu \textsc{Tominaga}\altaffilmark{1,4}, 
        Koji S. \textsc{Kawabata}\altaffilmark{6,7},
        Katsutoshi \textsc{Takaki}\altaffilmark{7},
        Miho \textsc{Kawabata}\altaffilmark{7},
        Tatsuya \textsc{Nakaoka}\altaffilmark{7},
        Issei \textsc{Ueno}\altaffilmark{7},
        Hiroshi \textsc{Akitaya}\altaffilmark{6,7}, 
        Takahiro \textsc{Nagayama}\altaffilmark{8,9}, 
        Jun \textsc{Takahashi}\altaffilmark{10}, 
        Satoshi \textsc{Honda}\altaffilmark{10}, 
        Toshihiro \textsc{Omodaka}\altaffilmark{9}, 
        Ryo \textsc{Miyanoshita}\altaffilmark{9},
        Takashi \textsc{Nagao}\altaffilmark{3},
        Makoto \textsc{Watanabe}\altaffilmark{11},
        Mizuki \textsc{Isogai}\altaffilmark{5,12},
        Akira \textsc{Arai}\altaffilmark{10,12},       
        Ryosuke \textsc{Itoh}\altaffilmark{7}, 
        Takahiro \textsc{Ui}\altaffilmark{7}, 
        Makoto \textsc{Uemura}\altaffilmark{6,7}
        Michitoshi \textsc{Yoshida}\altaffilmark{6,7},
        Hidekazu \textsc{Hanayama}\altaffilmark{13},
        Daisuke \textsc{Kuroda}\altaffilmark{14}, 
        Nobuharu \textsc{Ukita}\altaffilmark{14},
	Kenshi \textsc{Yanagisawa}\altaffilmark{14}, 
        Hideyuki \textsc{Izumiura}\altaffilmark{14},
        Yoshihiko \textsc{Saito}\altaffilmark{15},
        Kazunari \textsc{Masumoto}\altaffilmark{16},
        Rikako \textsc{Ono}\altaffilmark{16},
        Ryo \textsc{Noguchi}\altaffilmark{16},
        Katsura \textsc{Matsumoto}\altaffilmark{16},
        Daisaku \textsc{Nogami}\altaffilmark{2,3},
	Tomoki \textsc{Morokuma}\altaffilmark{17},
	Yumiko \textsc{Oasa}\altaffilmark{18}, and
        Kazuhiro \textsc{Sekiguchi}\altaffilmark{5}.}

\altaffiltext{1}{Department of Physics, Faculty of Science and Engineering, Konan University, Okamoto, Kobe, Hyogo 658-8501,
Japan; yamanaka@center.konan-u.ac.jp} 
\altaffiltext{2}{Kwasan Observatory, Kyoto University, 
17-1 Kitakazan-ohmine-cho, Yamashina-ku, Kyoto, 607-8471, Japan} 
\altaffiltext{3}{Department of Astronomy, Graduate School of Science, Kyoto University, Sakyo-ku, Kyoto 606-8502, Japan}
\altaffiltext{4}{Kavli Institute for the Physics and Mathematics of the
Universe (WPI), The University of Tokyo, 5-1-5 Kashiwanoha, Kashiwa, Chiba
277-8583, Japan}
\altaffiltext{5}{National Astronomical Observatory of Japan, Osawa, Mitaka, Tokyo 181-8588, Japan}
\altaffiltext{6}{Hiroshima Astrophysical Science Center, Hiroshima University, Higashi-Hiroshima, Hiroshima 739-8526, Japan}
\altaffiltext{7}{Department of Physical Science, Hiroshima University, Kagamiyama 1-3-1, Higashi-Hiroshima 739-8526, Japan} 
\altaffiltext{8}{Department of Astrophysics, Nagoya University, Chikusa-ku, Nagoya 464-8602, Japan}
\altaffiltext{9}{Graduate School of Science and Engineering, Kagoshima University, 1-21-35 Korimoto, Kagoshima 890-0065, Japan}
\altaffiltext{10}{Nishi-Harima Astronomical Observatory, Center for Astronomy, University of Hyogo, 407-2 Nishigaichi, Sayo-cho, Sayo, Hyogo 679-5313, Japan}
\altaffiltext{11}{Department of Earth and Planetary Sciences, School of Science, Hokkaido University, Kita-ku, Sapporo 060-0810, Japan}
\altaffiltext{12}{Koyama Astronomical Observatory, Kyoto Sangyo University, Motoyama, Kamigamo, Kita-Ku, Kyoto-City 603-8555, Japan}
\altaffiltext{13}{Ishigakijima Astronomical Observatory, National Astronomical Observatory of Japan, 1024-1 Arakawa, Ishigaki, Okinawa 907-0024, Japan}
\altaffiltext{14}{Okayama Astrophysical Observatory, National Astronomical Observatory of Japan, Honjo 3037-5, Kamogata, Asakuchi, Okayama 719-0232, Japan}
\altaffiltext{15}{Department of Physics, Tokyo Institute of Technology, 2-12-1 Ookayama, Meguro-ku, Tokyo 152-8551, Japan}
\altaffiltext{16}{Astronomical Institute, Osaka Kyoiku University, Asahigaoka, Kashiwara, Osaka 582-8582}
\altaffiltext{17}{Institute of Astronomy, Graduate School of Science, The University of Tokyo, 2-21-1 Osawa, Mitaka, Tokyo 181-0015, Japan}
\altaffiltext{18}{Faculty of Education, Saitama University, 255 Shimo-Okubo, Sakura, Saitama, 338-8570, Japan}



\maketitle
\begin{abstract}
 We present extensively dense observations of the super-Chandrasekhar 
supernova (SC SN) candidate SN 2012dn from $-11$ to $+140$ days after the date of its 
$B$-band maximum in the optical and near-infrared (NIR) wavelengths 
conducted through the OISTER ToO program. 
The NIR light curves and color evolutions up to 35 days after the $B$-band maximum 
provided an excellent match with those of another SC SN 2009dc, 
providing a further support to the nature of SN 2012dn as a SC SN.
We found that SN 2012dn exhibited strong excesses in the NIR wavelengths 
from $30$ days after the $B$-band 
maximum. The $H$ and $K_{s}$-band light curves exhibited much later 
maximum dates at $40$ and $70$ days after the $B$-band maximum, respectively, 
compared with those of normal SNe Ia.
 The $H$ and $K_{s}$-band light curves subtracted by those of SN 2009dc 
displayed plateaued evolutions, indicating a NIR echo from the 
surrounding dust. The distance to the inner boundary of the dust 
shell is limited to be $4.8 - 6.4\times10^{-2}$ pc.
No emission lines were found in its early phase spectrum, suggesting 
that the ejecta-CSM interaction could not occur.
On the other hand, we found no signature that strongly supports the
scenario of dust formation. The mass loss rate of the pre-explosion system 
is estimated to be $10^{-6}-10^{-5}$ M$_{\odot}$ yr$^{-1}$, assuming 
that the wind velocity of the system is $10-100$ km~s$^{-1}$,
which suggests that the progenitor of SN 2012dn could be a recurrent nova system. 
We conclude that the progenitor of this SC SN could be explained
by the single-degenerate scenario.

\end{abstract}


\section{Introduction}

  Type Ia supernovae (SNe Ia) have been used to provide constraints on 
 cosmological parameters through the use of their 
 calibrated light curves \citep{Riess1998,Perlmutter1999}. Strong correlations 
 exist between the rate of decline of the light curves and their 
 absolute magnitudes
 \citep{Phillips1993,Phillips1999,Altavilla2004,XWang2006,Prieto2006,Folatelli2010}.  
 Homogeneous properties are thought to be reproduced 
 by the similar masses of the progenitors (e.g., \cite{Branch1993}, 
 but see also \cite{Howell2006,Scalzo2010,Taubenberger2011} and therein).
 This would support a scenario that a thermonuclear runaway occurs at the 
 center or off-center region of a white dwarf (WD), when it reaches the 
 Chandrasekhar limiting mass ($\sim1.4$M$_{\odot}$) \citep{Nomoto1984}.

  However, recently various observational outliers have been discovered, 
 including overluminous SNe Ia (e.g., Super-Chandrasekhar SNe;
 \cite{Howell2006,Hicken2007,Yamanaka2009a,Tanaka2010a,Yuan2010,Scalzo2010,
 Silverman2011,Taubenberger2011,Scalzo2012,Chakradhari2014}), which represent
 a sub-class of SNe Ia studied in this paper. Super-Chandrasekhar SNe 
 (SC SNe) have peculiar observational
 properties, e.g., an extremely high luminosity, slow
 rate of decline of the optical light curves, typically very slow 
 expansion velocity, and anomalous carbon 
 absorption features \citep{Yamanaka2009a,Silverman2011,Taubenberger2011}. 
 The large amount of the ejected $^{56}$Ni mass, as inferred from 
 their peak luminosities, cannot be explained by the 
 Chandrasekhar-limiting mass of the WD \citep{Howell2006}. 
 Theoretical analysis of the light curves and spectra 
 imply that the total ejected mass could significantly exceed 
 the Chandrasekhar-limiting mass of a non-rotating WD and
 could even reach 2.0-3.0$M_{\odot}$ 
 \citep{Taubenberger2011,Hachinger2012,Kamiya2012}, 
 although an asymmetricity of the ejecta without invoking the 
 large $^{56}$Ni mass has also been discussed
 (\cite{Hillebrandt2007}, but see \cite{Maeda2009b,Tanaka2010a}). 


  Demonstrating the presence of circumstellar material (CSM) is 
 critical for providing a 
 constraint on the progenitor nature of an exploding star. 
 The dense environment of the CSM could be formed by 
 the shell being 
 ejected through the mass loss from the companion star 
 or through an optically thick 
 wind arising from the WD. This is expected from a 
 single-degenerate scenario \citep{Nomoto1982}, while the environment 
 surrounding two WDs would be scarce in a 
 double-degenerate scenario \citep{Iben1984,Webbink1984}. 
 Various possibilities for the existence of the CSM have been 
 discussed, based on light echoes in the 
 late-phase light curve mostly in the optical wavelength 
 \citep{Schaefer1987,Sparks1999,Cappellaro2001,
 Patat2005,LWang2005,Patat2006,Quinn2006,XWang2008b,Crotts2015,
 Drozdov2015} and 
 time-variable narrow absorption line systems for some SNe Ia 
 \citep{Patat2007,Simon2009}.
 The stronger evidence of CSM has been reported for a class of SNe Ia-CSM 
 showing a clear signature of the hydrodynamical interaction between the SN ejecta 
 and the CSM \citep{Hamuy2003,Dilday2012}. We note that 
 another important clue, a non-degenerate companion star, has been generally 
 non-detected for normal SNe Ia either from the direct detection in pre-SN 
 images \citep{Li2011,Kelly2014} or the
 shock-heated early emission \citep{Nugent2011,Yamanaka2014}, while the 
 possible signature has been reported for SN 2012cg \citep{Marion2016}. 
 These various methods for 
 detecting a companion star or CSM have not been applied to SC SNe to date.

  Thermal emission from CS dust
 radiated by light originating from the central SN can also be detected 
 in near-infrared (NIR) wavelength observations \citep{Maeda2015} if CS dust 
 is present.
 However, the geometrical structure of the progenitor system may affect the 
 ability to detect such signatures. In any case, if the 
 signatures of the CSM are acquired on subparsec scales, 
 the nature of the progenitor should be limited to the single degenerate state. 
 
  SN 2012dn was discovered at a magnitude of 16.3 on Jul 13.3 (UT)
 \footnote{We adopted $t=0$ as MJD 56132.39 derived by 
 \citet{Chakradhari2014}} in the 
 nearby galaxy, ESO 462-16 ($\mu=33.15$; \cite{Bock2012}). Its spectrum 
 resembled that of SC
 SN 2006gz at 10 days before maximum \citep{Parrent2012}. 
 \citet{Chakradhari2014} noted that the 
 optical 
 and ultraviolet properties of SN 2012dn are similar 
 to those of 
 SC candidate SN 2006gz. 
 Therefore, SN 2012dn could be the nearest SC SN Ia candidate to date.
 \citet{Parrent2016} performed the detailed spectral analysis of SN 2012dn.
 They derived the ejected mass exceeding $1.6 M_{\odot}$,
 indicating that SN 2012dn is a SC SN Ia.
 They also suggest that the origin of SN 2012dn might not be a merger event.


  In this paper, we present intensitively obtained samples of the optical 
 and NIR observations of the super-Chandrasekhar 
 candidate SN 2012dn obtained through the Target-of-Opportunity (ToO) program in 
 the Optical and Infrared Synergetic Telescopes for Education and Research (OISTER). 
 We focus on the NIR data, which have never before been 
 published. The NIR light curves and color exhibited 
 strong excesses after $t=30$ d. The $J$, $H$, and $K_{s}$-band 
 fluxes were subtracted using the light curves of SN2009dc as 
 a template. We demonstrate that the subtracted light 
 curves could be naturally explained by dust echo models. 
 The classification of SN 2012dn as a SC SN is also discussed using 
 the NIR data. The nature of the progenitor system 
 is then discussed using the mass loss rate and the distances
 to the CS dust shell derived from the dust echo model.
 Finally, we present our conclusion that a SC SN progenitor could 
 originate from the single-degenerate scenario.

\begin{figure*}
  \begin{center}
    \begin{tabular}{c}
      \resizebox{160mm}{!}{\includegraphics
 {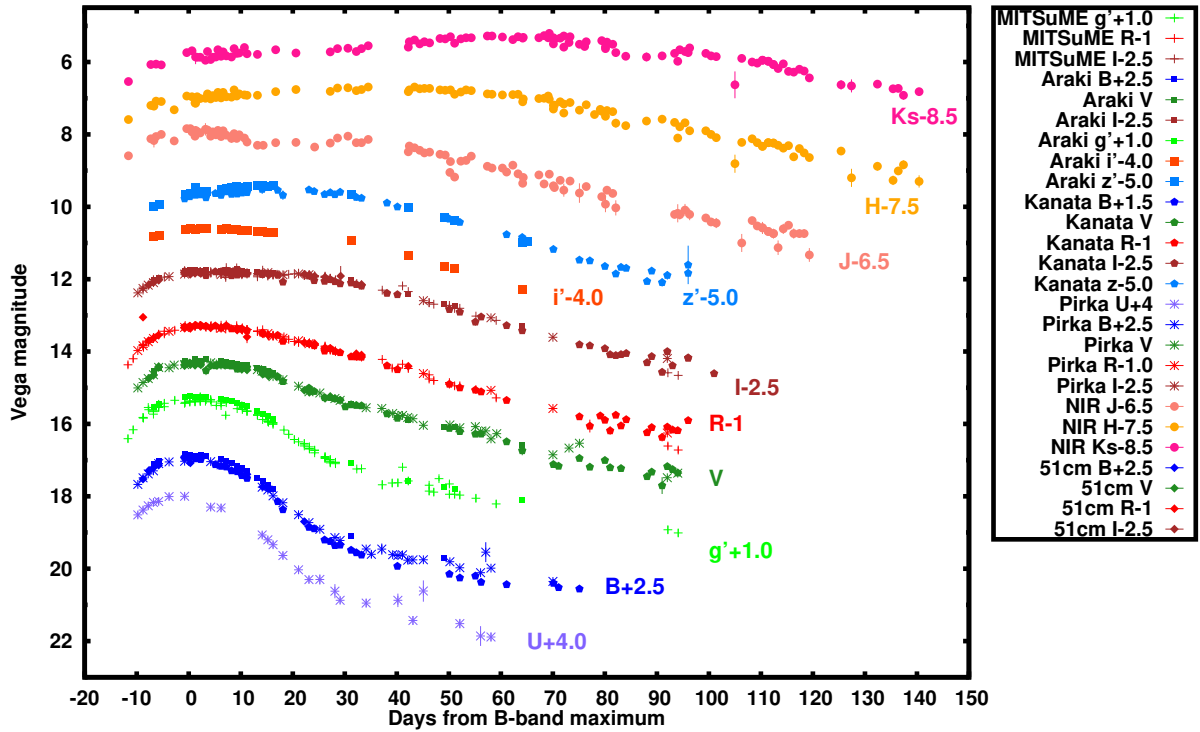}} \\
    \end{tabular}
  \end{center}
    \caption{$U$, $B$, $g'$, $V$, $R$, $I$, $i'$, $z'+Y$, $J$, $H$
and $K_{\rm s}$-band light curves of SN 2012dn. The $U$, $B$, $V$, $R$
 and $I$-band light curves are consistent with those presented in 
 \citet{Chakradhari2014}. Corrections have been made for 
 Galactic and host galactic extinctions. The legend denotes the 
 photometric data obtained by each instrument. The detailed observational 
 modes are listed in Table 1.} 
    \label{epl}
\end{figure*}

\section{Observations and data reduction}


 Using the OISTER 
 (\cite{Itoh2013,Itoh2014,Ishiguro2015,Kuroda2015,Yatsu2015,Yamanaka2015}), 
 ToO program, observations were 
 conducted for SN 2012dn between $t=-7$ and $140$ d.
 Optical and NIR photometric observations of SN 2012dn were 
 performed using different 10 telescopes, while 
 spectroscopic observations were carried out using one telescope.

 \subsection{Optical photometry}

 We performed $B$, $V$, 
 $R$, $I$, and $z'+Y$-band photometric observations over 52 nights 
 using the Hiroshima One-shot Wide-field Polarimeter 
 (HOWPol; \cite{Kawabata2008}) 
 installed on the 1.5-m Kanata telescope,  
 $U$, $B$, $V$, $R$, and $I$-band observations on 44 nights 
 with the Multi-spectral Imager (MSI; \cite{Watanabe2012}) 
 installed on the 1.6-m Pirka 
 telescope, and $B$, $V$, $I$, $g'$, $i'$, $z'$-band 
 photometric observations over 24 nights with the Araki telescope 
 DuaL-band imagER (Adler; \cite{Isogai2015}) installed on the 
 1.3-m Araki telescope. The observatories, telescopes, and 
 instruments participating in this program are summarized in Table 1.

  We also performed $g'$, $R$, and $I$-band 
 observations using a robotic observation system, the 
 Multicolor Imaging Telescopes for Survey and Monstrous Explosions 
 (MITSuME; \cite{Kotani2005}), at the Okayama Astrophysical Observatory
 (OAO) \citep{Yanagisawa2010} on 67 nights; at the Akeno Observatory on 
 11 nights; and at the Ishigaki-jima Astronomical Observatory on 
 29 nights. Expanding the OISTER collaboration, we performed $B$, 
 $V$, $R$, and $I$-band photometric observations on 29 nights using the 
 0.5-m (51-cm) telescope at Osaka Kyoiku University.
 The Kyoto Okayama Optical Low-dispersion Spectrograph 
 (KOOLS; \cite{Yoshida2005}) attached to the Cassegrain 
 focus of the 1.88-m OAO telescope was used for observations 
 of the photometric standard stars on one night. 
 These data were reduced using a standard procedure for 
 charge-coupled device (CCD) photometry. 
 The obtained data were reduced 
 according to a standard procedure for ground-based observations.
 We performed point spread 
 function (PSF) fitting photometry using the DAOPHOT 
 package of the IRAF \footnote{IRAF is distributed by the National Optical 
 Astronomy Observatories, operated by the Association of Universities for 
 Research in Astronomy, Inc., under contract to the National Science 
 Foundation of the United States.} reduction and analysis software.

  \subsection{Near-infrared photometry}
 $J$, $H$, and $K_{\rm s}$-band photometric observations were 
 performed over 80 nights using the Simultaneous three-color InfraRed Imager
 for Unbiased Survey (SIRIUS; \cite{Nagayama2003}) installed on 
 the 1.4-m InfraRed Survey Facility (IRSF) telescopes at the 
 South Africa Astronomical Observatory (SAAO); 
 the Nishi-harima Infrared Camera (NIC) installed at the 
 Cassegrain focus of the 2.0-m Nayuta telescope at the 
 Nishi-Harima Astronomical Observatory on 21 nights;  
 and the InfraRed Camera installed on the 1.0-m telescope at 
 the Iriki Observatory on 24 nights.
 The sky background was subtracted using a template sky 
 image derived from each dithering observation set. PSF fitting 
 photometry was conducted in a similar manner to the
 optical.

 \subsection{Photometric calibrations}
  Photometric calibrations were performed using a similar 
 method to \citet{Yamanaka2015}.
  Relative photometry was carried out using the local reference
 stars in the field of SN 2012dn. For the NIR data,
 the 2MASS catalog magnitudes were used for the
 reference stars \citep{Persson1998}. A square average of the 
 standard deviation of the SN and the systematic
 error of the reference star magnitudes were adopted 
 for the observational errors. 
 A summary of the results of the photometry is listed in Table $1-7$.
 Photometric data for standard stars
 in the SA 111, SA 112, and SA 113 regions \citep{Landolt1992}, which 
 were obtained using HOWPol, MSI, Adler, and KOOLS on photometric observation nights,
 were used to calibrate the $U$, $B$, $V$, $R$, $I$, $g'$, $i'$, and $z$-band 
 magnitudes of the reference stars. Using these magnitudes, 
 relative photometry for the SN was performed.
 The calibrated reference magnitudes were consistent with 
 those of \citet{Chakradhari2014}, which lay within the error.
 Note that we reduced the systematic error
 for the various instruments using a color-term correction in
 a consistent manner and included newly performed observations
 of the standard stars in the M67 \citep{Stetson1987} and SA98
 \citep{Landolt1992} regions.

 \subsection{Optical spectroscopy}
   We performed optical spectroscopic monitoring using
 HOWPol over a period of 30 nights.
 The wavelength coverage was $4,500-9,000$ \AA\
 and the
 spectral resolution was R=$\lambda$/$\Delta$$\lambda$ $\sim$ 
 400 at 6,000 \AA. 
 During the data reduction procedure, the wavelengths were calibrated using
 telluric emission lines acquired in the object frames.
 The fluxes were calibrated using the frames of
 spectrophotometric standard stars taken on the same night as
 the object data. We removed strong telluric absorption
 features from the spectra using the spectra of high-temperature standard stars.

\section{Observational properties} 

\subsection{Light curves}
 Figure 1 shows the multi-band optical and NIR photometric 
data for SN 2012dn, where the magnitudes have already been corrected 
for Galactic and host galactic extinctions \citep{Schlafly2011,Chakradhari2014}. 
$U$, $B$, $V$, $R$, and $I$-band data were consistent with 
those reported by \citet{Chakradhari2014}, which lay within our error.
The NIR light curves exhibited a single-peaked 
maximum, except for the $J$-band. Redder-band light curves 
arrived at their maximum at a later epoch.
The $H$ and $K_{\rm s}$-band maxima were particularly
visible at $\sim40$ and $\sim70$ days after the $B$-band maximum date, respectively.
Such large delays have never been found in any SNe Ia to date. 


\begin{figure}
  \begin{center}
    \begin{tabular}{c}
      \resizebox{90mm}{!}{\includegraphics{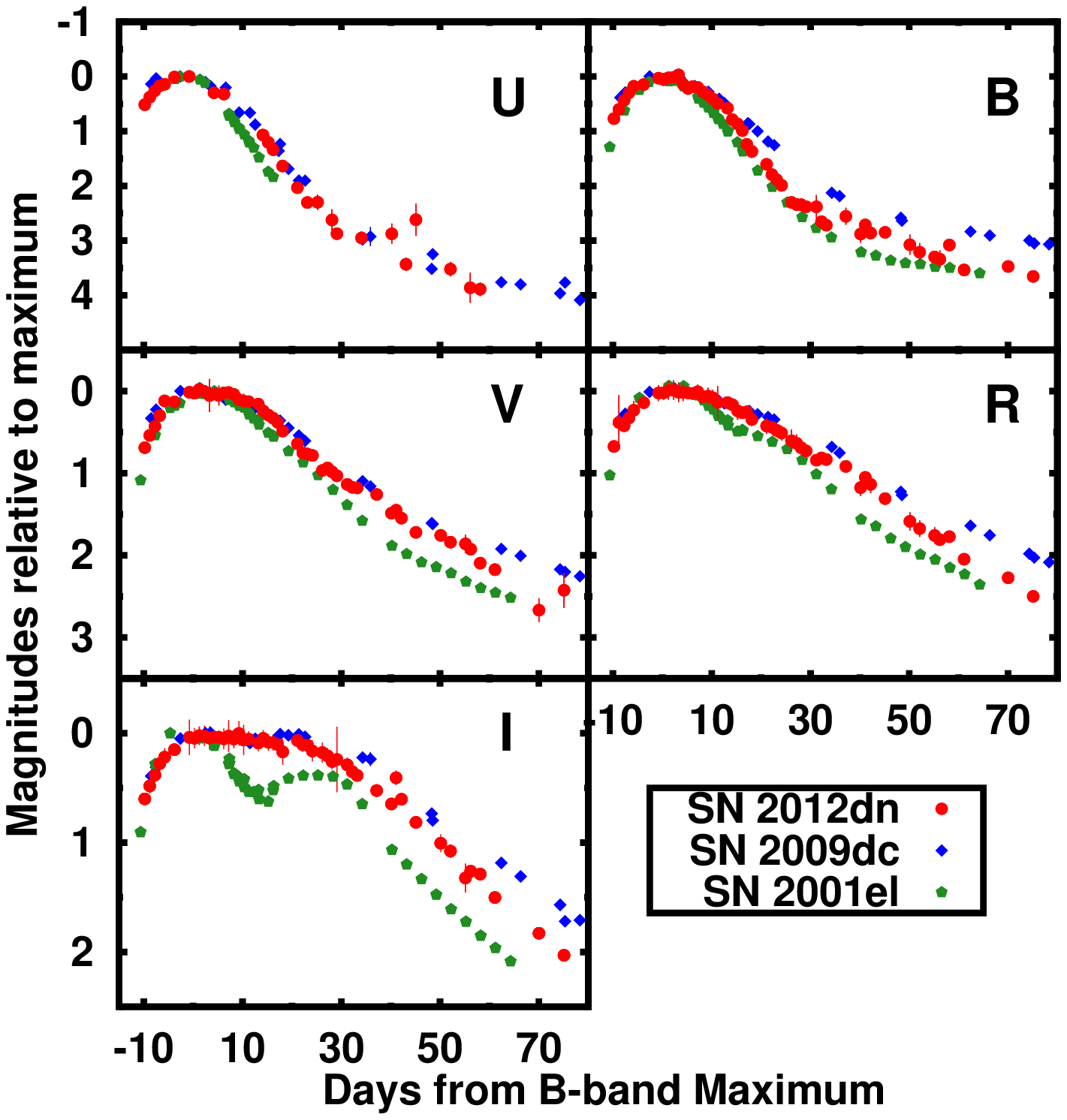}} \\
    \end{tabular}
 \end{center}
    \caption{$UBVRI$-band light curves of SN 2012dn compared to 
 another well-observed SC SN 2009dc \citep{Yamanaka2009a,Silverman2011,Taubenberger2011} 
 and a typical SN Ia 2001el \citep{Krisciunas2004}. Our data
 are denoted by the red-filled circles. Those of SNe 2001el and 2009dc are denoted by the 
 blue diamond and green pentagons, respectively. The magnitudes were 
 artificially shifted to their maximum for comparison. The data were stacked into the 
 averaged magnitudes within a day from our data.} 
    \label{epl}
\end{figure}

\begin{figure*}
  \begin{center}
    \begin{tabular}{c}
      \resizebox{170mm}{!}{\includegraphics{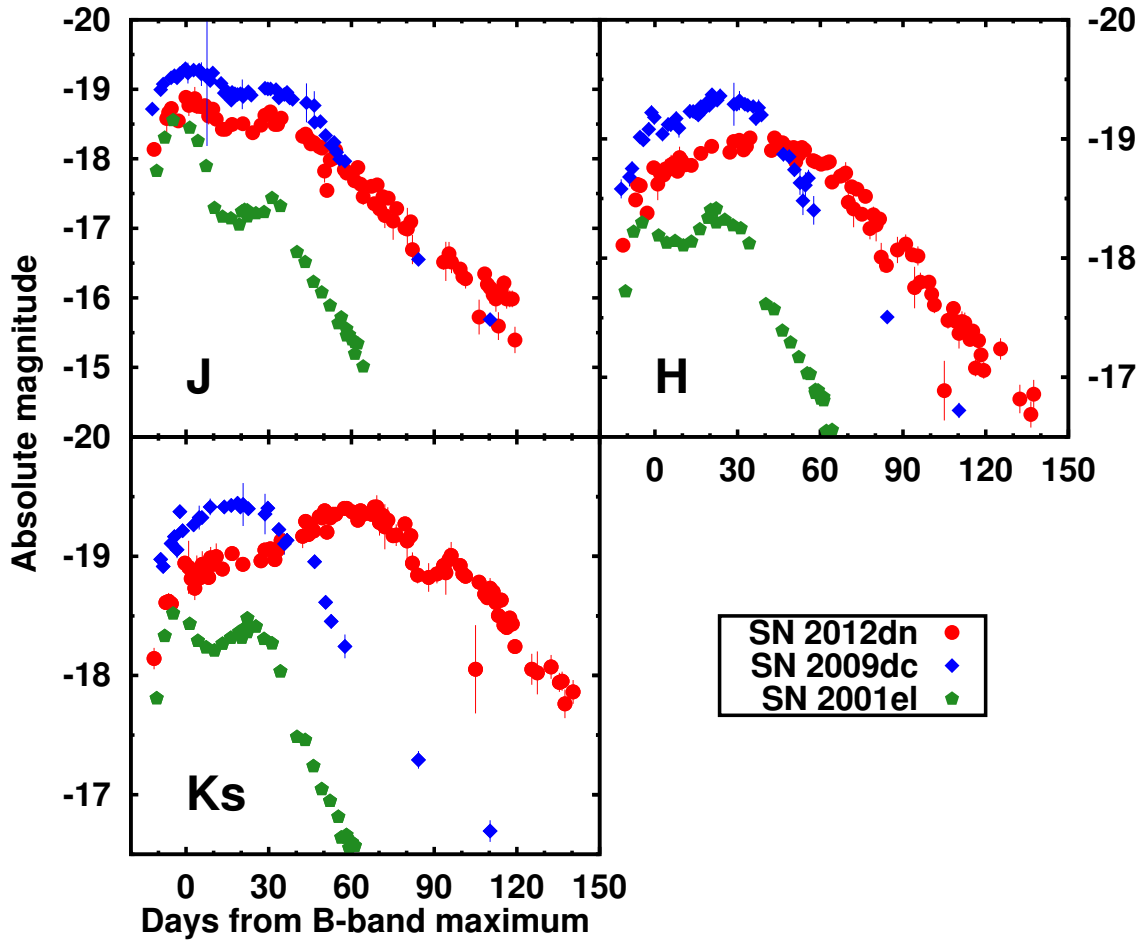}} \\
    \end{tabular}
  \end{center}
    \caption{$JHK_{\rm s}$-band absolute magnitude light curves of SN 2012dn 
    compared with those of SNe 2001el and 2009dc. Additional data for SN 2009dc 
    from \citet{Friedman2015} were included for comparison. The absolute magnitudes of
    SN 2012dn were calculated using a distance modulus of $\mu=33.15$ to the 
    host galaxy, ESO 462-16. SNe are denoted in a similar 
    manner to Fig. 2. A dip at $\sim 80 - 90$ days seen in $H$ and $K_{s}$ bands (and possibly
in the $J$ band) is probably a real feature, since the
data independently obtained using different three telescopes were
consistent within the systematic error. We stacked these data taken
within a day into the averaged magnitudes.
This feature would not contradict to the general 
interpretations for the NIR excess we examined in this paper, but this may contain 
information to further constrain on details of each scenario. 
We postpone such a study 
to the future.}
    \label{epl}
\end{figure*}

 Figure 2 provides a comparison of the optical $U$, $B$, $V$, $R$,  
and $I$-band light curves with those of SNe 2001el \citep{Krisciunas2004}
and 2009dc \citep{Yamanaka2009a,Silverman2011,Taubenberger2011}. 
The $U$-band data provide better coverage than those presented by \citet{Chakradhari2014}.
The rising part of the $U$-band light curve exhibited a slightly slower 
evolution than that of typical SN 2001el, while it was similar to 
that of another SC SN 2009dc \citep{Taubenberger2011}.
The $U$-band maximum date was consistent with those of SNe 2001el and 
2009dc. The rates of decline of the light curves for all bands 
were much slower than those of SN 2001el. The $B$-band light 
curve showed a significantly faster evolution than that of
another SC SN 2009dc. The decline rates of the $U$, $V$, $R$ and $I$-band 
light curves were relatively similar to those of SN 2009dc.

  Figure 3 provides a comparison of the absolute magnitude light curves in the 
 $J$, $H$, and $K_{\rm s}$ bands of SN 2012dn with those of SNe 2001el and 2009dc.
 Additional data for SN 2009dc from \citet{Friedman2015} were included for this comparison. 
 The light curves shown in Fig. 3 are the average of those of obtained within a night. 
 The $J$-band light curve
 of SN 2012dn exhibited first and secondary maxima.
 The first peak was reached a few days after the $B$-band maximum, while for
 typical SN 2001el the first peak was reached a few days before the maximum.
 The secondary peak occurred at $t=35$ d, which was similar to that 
 of typical SN 2001el. The magnitude of the $J$-band secondary maximum 
 was $\sim0.2$ mag fainter than the first, while that of SN 2001el
 was $\sim1.5$ mag fainter. The overall shape
 was fairly similar to that of SN 2009dc up to $t=40$ d, but the decline rate 
 after $t=40$ d was much lower than that of both SNe.
 

 For the $H$ band, the magnitudes reached the maximum at around 
$t=40-50$ d.
The light curve almost linearly declined from $t=40$ d until $t=140$. 
The rate of decline was estimated to be 2.5 mag
per 100 d between $t=40$ and $120$ d, which was much slower than the
3.8 mag per 100 d between $t=20$ and $60$ d for the typical SN 2001el.
The shape of the light curve was comparable to that 
of SN 2009dc up to $t=20$ d, but the light curve was less luminous
up to $t=40$ d. SN 2012dn had a much 
more luminous absolute magnitude than SN 2009dc after $t=45$ d, and the decline rate was 
much slower between $t=45$ and $110$ d. 

 For the $K_{s}$ band, the above trend became even more noticeable. The light 
curve increased rapidly until $t=0$ d and then slowly evolved 
between $t=0$ and $40$ d. The light curve reached 
its maximum at $t=60-70$ d. The decline rate 
was estimated to be 2.0 mag per 100 days between $t=70$ and $140$ d,
indicating that it had the slowest decline among the $U$ to $K_{s}$ bands.
The shape of the light curve of SN 2012dn was similar to that 
of SN 2009dc up to $t=25$ d. SN 2012dn became fainter
than SN 2009dc after $t=40$ d, while its magnitude was 1.5 times more luminous 
than SN 2009dc between $t=85$ and $110$ d. 



\begin{figure*}
 \begin{center}
    \begin{tabular}{c}
      \resizebox{85mm}{!}{\includegraphics{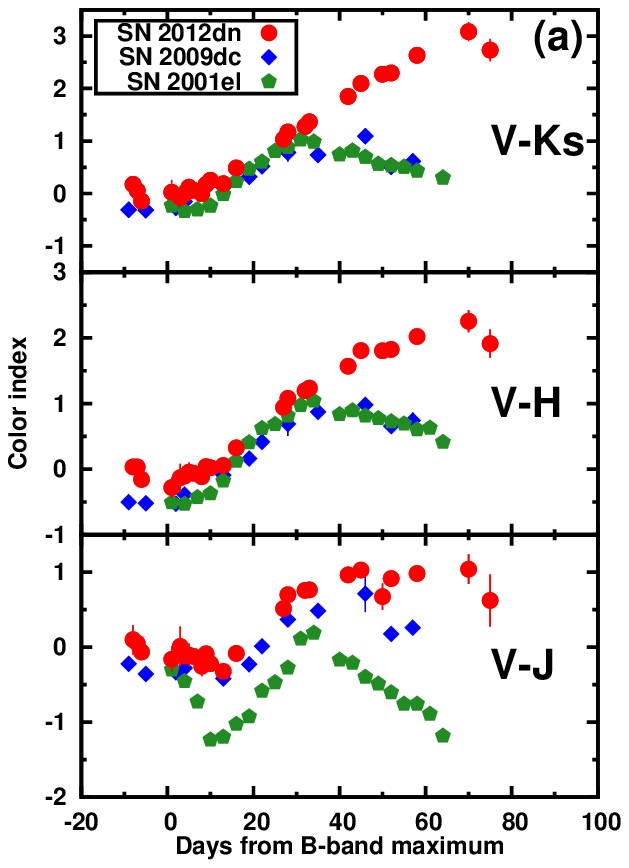}} 
      \resizebox{85mm}{!}{\includegraphics{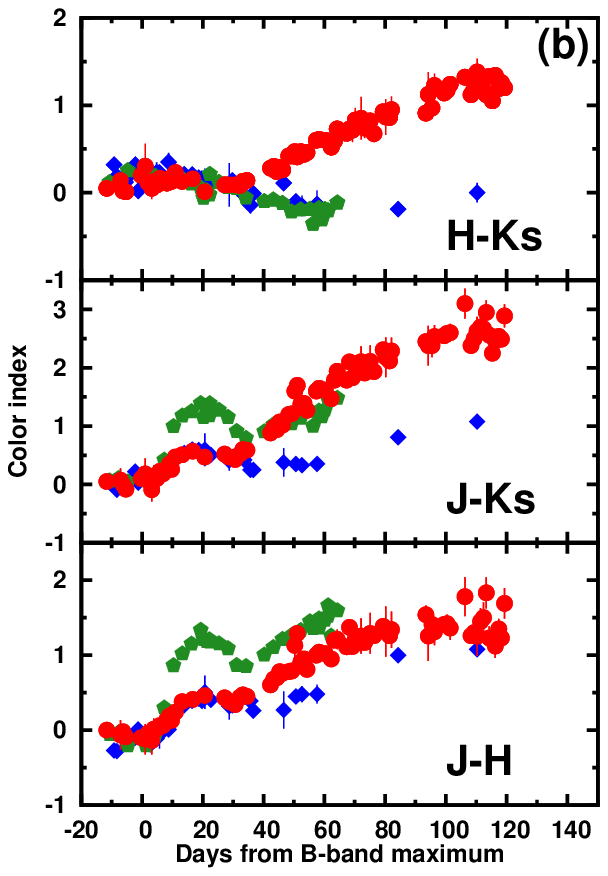}} \\
    \end{tabular}
  \end{center}
    \caption{(a) $V-J$, $V-H$, and $V-K_{\rm s}$ color evolutions of SN 2012dn 
    compared with those of SNe 2001el and 2009dc. The legends are the same 
    as given in Fig. 2. (b) $J-H$, $J-K_{\rm s}$, and $H-K_{\rm s}$ color 
    evolutions compared with those of SNe 2001el and 2009dc. Symbols are denoted
    in the same manner as in Fig. 2.} 
    \label{epl}
\end{figure*}

\subsection{Color evolutions}
 Figure 4 provides a comparison of the $V-J$, $V-H$, and $V-K_{\rm s}$ 
color evolutions with those of SNe 2001el and 2009dc. The $V-J$ color
exhibited a slow evolution to a much bluer color until $t=15$ d. 
Thereafter, the color evolved 
to become redder until $t=30$ d, becoming flat between $t=30$ and $100$
d. The $V-J$ color of SN 2012dn was always redder than that of 
SN 2001el, but was similar to that of SN 2009dc until
$t=40$ d while reddening after $t=40$ d.

The $V-H$ and $V-K_{s}$ color evolutions are notable. 
The $V-H$ color exhibited a monotonous evolution to a much redder color until 
$t=40$ d. Thereafter, the evolution slowed slightly until $t=70$ d. 
The evolution up to $t=40$ d was similar to that of SNe 2001el and 
2009dc, while the color became much redder than both SNe after $t=40$ d.
Similar trends were confirmed for $V-K_{s}$, with the evolution up to $t=40$ d
being similar to that of both SNe. 
The color was approximately $\sim3.0$ mag redder than that of both SNe at $t=60-70$ d. 



 Figure 4 also shows the $J-H$, $J-K_{s}$, and $H-K_{s}$ color evolutions compared 
with those of SNe 2001el and 2009dc. The color evolutions exhibited
rather interesting behavior. The $J-H$
color of SN 2012dn was always bluer than that of SN 2001el after $t=10$ d. 
The color began to evolve to be bluer up to $t=30$ d, 
similar to that of SN 2001el. 
The color evolution was also almost as fast as that of SN 2001el
after $t=30$ d, while the color was bluer. In comparison with SN 2009dc, 
the color was similar to that until $t=35$ d, after which it became redder. 

 Similar trends were observed in the $J-K_{s}$ color evolution. 
The break points at $t=30$ d came slightly earlier 
than those of SN 2001el, and the evolution slope was marginally bigger
after the $t=30$ d period. The color became redder 
than that of SN 2001el after $t=50$ d, but the color evolution was similar to 
that of SN 2009dc up to $t=30$ d, becoming much redder than that of 
SN 2009dc at $t=35$ d.

 The $H-K_{s}$ color evolution was similar to that of SNe 2001el and 
2009dc until $t=30$ d. Thereafter, the color evolved 
to become redder while those of SNe 2001el and 2009dc became bluer. 
The deviation of the $H-K_{s}$ color evolution
of SN 2012dn implies that there is an external component in the NIR region. 
The subtraction of the NIR light curves by those of SN 2009dc is carried out
in \S 5.1 to clarify the nature of this external component.

\begin{figure*}
  \begin{center}
    \begin{tabular}{c}
      \resizebox{120mm}{!}{\includegraphics{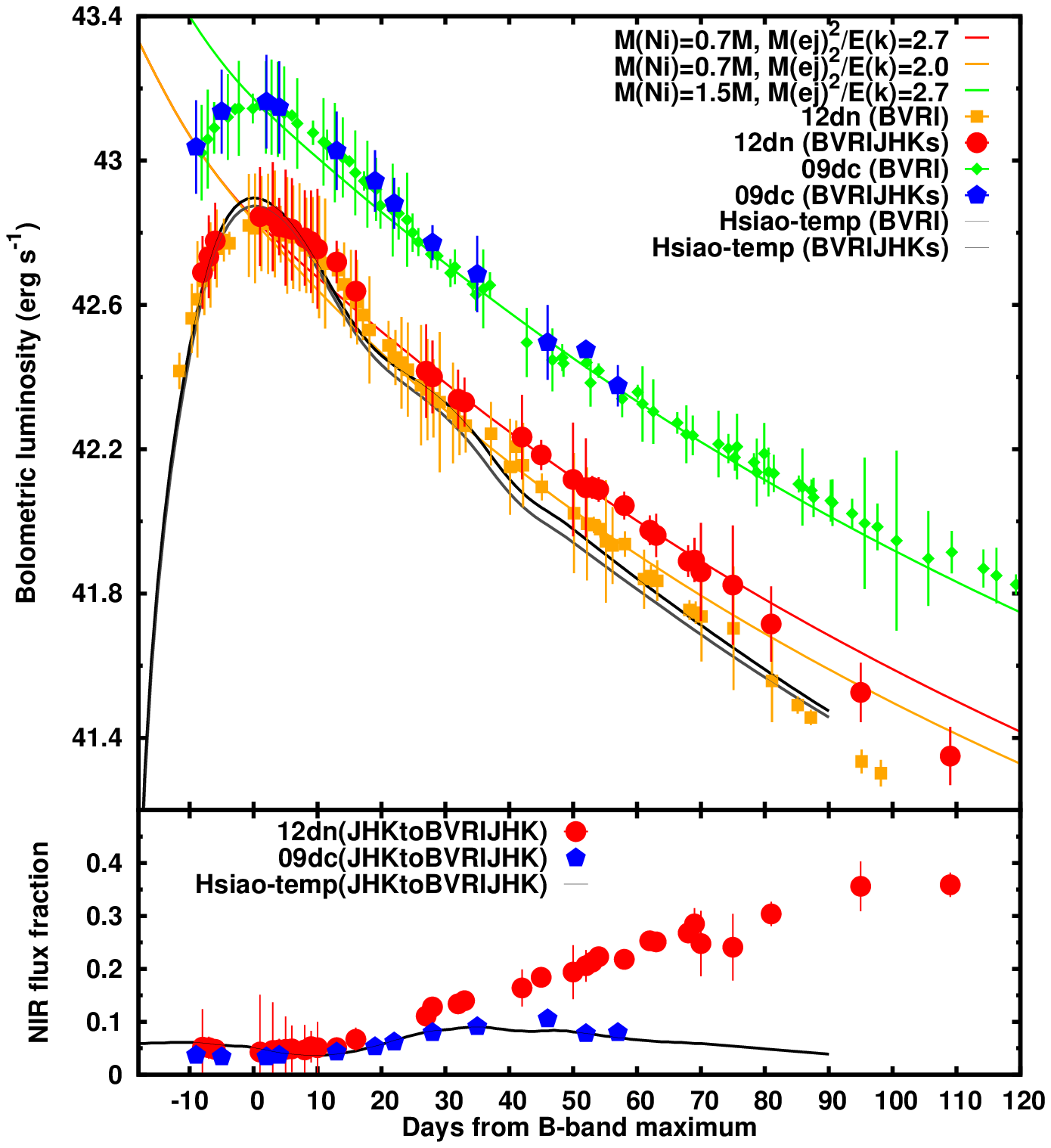}} \\

    \end{tabular}
  \end{center}
    \caption{(Top panel) Quasi-bolometric light curves of SN 2012dn integrated with 
$BVRI$ and $BVRIJHKs$ -band photometric data, which are denoted by
the orange-filled squares and red-filled circles, respectively. The light curves
of SC SN 2009dc \citep{Yamanaka2009a,Silverman2011,Taubenberger2011} and 
 the SN Ia template \citep{Hsiao2007} are plotted for comparison. 
 The analytical 
light curves with ($M/M_{\odot}^{2}$)$\cdot$($E_{K}/10^{51}$ erg~s$^{-1}$)$^{-1}=2.0$ 
and $2.7$ are also plotted. The $^{56}$Ni mass was 0.7$M_{\odot}$ for both analytical 
 curves. They are denoted by the orange and red lines, respectively.
The analytical light curve for SN 2009dc with 
($M/M_{\odot}^{2}$)$\cdot$($E_{K}/10^{51}$ erg~s$^{-1}$)$^{-1}=2.7$ is also plotted.
(Bottom panel) Flux fraction evolutions of the $JHKs$ to $BVRIJHK_{s}$-band 
integration flux of SNe 2012dn and 2009dc. 
That of the SN Ia template is also plotted and is denoted by the black line. Symbols are the same as for the upper panel.} 
    \label{epl}
\end{figure*}

 \subsection{Bolometric luminosity}
 The $BVRI$ and $BVRIJHK_{\rm s}$ -band integration light curves 
were constructed using the full width half maximum (FWHM) 
and the center wavelengths of 
their passband \citep{Fukugita1996}. Galactic and host 
galactic extinctions were corrected using 
$E$($B-V$)$=0.1$ \citep{Chakradhari2014}. 
A distance modulus of
$\mu=33.15$ was adopted \citep{Theureau1998}. 
Additional data from \citet{Chakradhari2014} were also included.
The optical-NIR integration light curve was compared with 
the optical alone to assess whether the NIR excess originated from a 
different component from the SN emission.
For comparison, the $BVRIJHK_{s}$ and $BVRI$ -band integration
light curves of another SC SN 2009dc and the template SN Ia 
were also constructed \citep{Hsiao2007} (see Fig. 5).
The absolute magnitudes of the template were obtained from the
decline rate for a normal SN Ia 
($\Delta$m$_{15}$($B$)$=1.1$; \cite{Phillips1999})

The peak luminosity was estimated to be $7.0\times10^{42}$ erg~s$^{-1}$. 
Assuming that the luminosity was $60\%$ in the $BVRIJHK_{s}$-band 
coverage region \citep{Stritzinger2006}, the quasi-bolometric luminosity was
calculated to be $1.2\times10^{43}$erg~s$^{-1}$. 
The luminosity was $25\%$ fainter than the luminosity estimated using 
a similar method by \citet{Chakradhari2014}. 
This difference could be caused by a systematic error. 
The peak luminosity was slightly fainter than that of the 
template light curve, indicating that the absolute luminosity of
SN 2012dn was similar to normal SNe Ia. The peak luminosity was 
2.3 times fainter than that of the extremely luminous SN 2009dc.

 There was a slight difference in the decline of the light curve 
between the optical and optical-NIR integrations.
 The decline rate of the $BVRIJHK_{s}$-band light curve
between $t=0$ and $75$ d was estimated to be 
8.5$\times10^{40}$ erg~s$^{-1}$~d$^{-1}$, which was 
marginally slower than the $8.7 \times10^{40}$ erg~s$^{-1}$~d$^{-1}$ 
for the $BVRI$-band integration. The difference increased
at later epochs. The decline rates of the $BVRIJHK_{s}$ and 
$BVRI$-band integration light curves of SN 2009dc 
were comparable.

 The both light curves exhibited the faster decline than 
 the analytical models after $t=80$ d. Later-epoch light curve for an another 
 SC SN 2006gz also exhibited faster decline 
 \citep{Maeda2009b,Chakradhari2014}.
 Those of SNe 2007if and 2009dc
 were quite slower \citep{Chakradhari2014}. There could be diversity of 
 the light curve decline among SC SNe. Thus, the faster decline of
 the light curve of SN 2012dn could be intrinsic. On the other hand, the 
 NIR light curves exhibited slower evolution after $t=80$ d, indicating
 the echo component (see \S 5.2.2), but the energy contributions were
 relatively small.

The evolution of the NIR contribution was then investigated. 
The lower panel of Fig. 5 
shows the time evolution of the NIR flux fraction in the 
optical-NIR integration light curve.
The NIR flux dramatically increased
after $t=15$ d. This was earlier than our expectation based
on a comparison of the color evolutions with those of SNe 2001el
and 2009dc (see \S 3.2). The NIR flux contribution became 
35 \% of the total optical-NIR integration flux at $t=110$ d.
On the other hand, the NIR contributions of normal SNe 
Ia and 2009dc were lower than those at earlier epochs 
\citep{Sollerman2004,XWang2009a}.
This was a unique feature of SN 2012dn.
 


\begin{figure}
  \begin{center}
    \begin{tabular}{c}
      \resizebox{85mm}{!}{\includegraphics{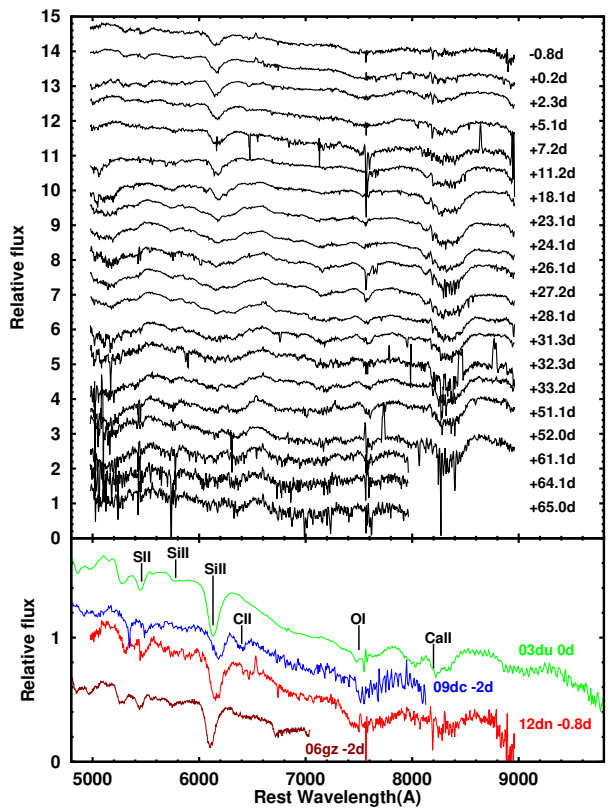}} \\
    \end{tabular}
  \end{center}
    \caption{(Top panel)
Time series of the spectra of SN 2012dn obtained by HOWPol 
between $t=-0.8$ and $65.0$ d. The wavelengths were converted to the rest frame 
using the recession velocity of the host galaxy, ESO 462-16. The atmospheric 
absorption lines were corrected using the spectra of standard stars obtained 
on the same night as the objects. (Bottom panel) The spectrum at the maximum compared
to those of SNe 2003du \citep{Stanishev2007}, 2006gz \citep{Hicken2007} and
 2009dc \citep{Yamanaka2009a}. The data for SNe 2003du and 2006gz were taken from
 the SUSPECT database \protect \footnotemark.} 
    \label{epl}
\end{figure}
\footnotetext{http://bruford.nhn.ou.edu/~suspect/index.html}

\begin{figure}
  \begin{center}
    \begin{tabular}{c}
      \resizebox{85mm}{!}{\includegraphics{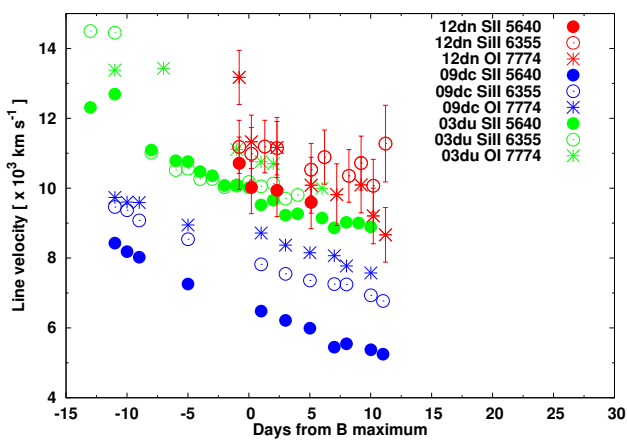}} \\
    \end{tabular}
  \end{center}
    \caption{
Line velocity evolutions of the O~{\sc i}$\lambda$7774, Si{\sc ii}$\lambda$6355, 
and S~{\sc ii}$\lambda$5640 absorption lines of SNe 2009dc and 2012dn.
The data for SNe 2003du and 2009dc were taken from the SUSPECT and 
WISEREP \protect \footnotemark databases.
Measurements of the absorption lines were performed using the $splot$ command 
in $IRAF$. 
The error was calculated by summing the standard deviations and wavelength resolution.} 
    \label{epl}
\end{figure}
\footnotetext{http://wiserep.weizmann.ac.il/}

 \subsection{Spectral properties}
 
  The spectral evolution of SN 2012dn is shown in Fig. 6. 
 The wavelengths of the data were corrected using the recession
 velocity of the host galaxy \citep{Theureau1998}. The spectra exhibited
 the multiplet of the Fe~{\sc ii}, Si~{\sc ii}, S~{\sc ii} W-shape
 features, the Ca~{\sc ii} IR triplet, and the O~{\sc i}, 
 indicating that this SN should be classified into the subclass of SNe Ia 
 \citep{Branch1993}. The strong C~{\sc ii}~$\lambda$6580 and $\lambda$7234
 absorption lines were detected until $t=25$ d, which 
 was 10 days later than the last one reported by \citet{Chakradhari2014},
 although these may be contaminated by Fe~{\sc ii} 
 absorption during the later phase because emission lines from 
 iron-group elements are dominant in the inner-ejecta region.
 The C~{\sc ii} absorption lines after the $B$-band maximum 
 are also confirmed in SN 2009dc \citep{Yamanaka2009a}.

 We measured the absorption minima, O~{\sc i}~$\lambda$7774, 
 Si~{\sc ii}~$\lambda$6355, and S~{\sc ii}~$\lambda$5640.
 The line velocities of these features of the normal SN 2003du 
 \citep{Stanishev2007} and SC SN 2009dc were also measured in the same manner.
 The measured features were identified as indicated in the lower panel of
 Fig. 6 by comparison with those of well-studied SC SNe 
 \citep{Hicken2007,Yamanaka2009a}.
 The line velocity evolutions are also shown in Fig. 7. 
 The decline rate of the line velocity evolution of S~{\sc ii} was 
 the largest among these features. Similar trends were also detected in 
 SNe 2003du and 2009dc.
 The Si~{\sc ii} line velocity exhibited a moderate decline,
 which was similar to those of SNe 2003du and 2009dc. In contrast to 
 the decline rates of Si~{\sc ii} and S~{\sc ii}, the decline rate of 
 the O~{\sc i} line velocity was much larger than that of 
 SN 2009dc. Finally, the velocities of all the elements were $1.3-1.7$ times 
 faster than those of SN 2009dc, but similar to those of SN 2003du 
 near the $B$-band maximum.

  \section{Classification as a super-Chandrasekhar SN Ia}

 \citet{Chakradhari2014} reported 
 that the spectral properties displayed a resemblance to 
 those of the SC SN candidate SN 2006gz. The strong C~{\sc ii} and
 shallow Si~{\sc ii} absorption lines were similar to those of 
 other SC SNe, and the line velocities were comparable to those of SN 2006gz, 
 rather than the SNe 2003fg and 2009dc. They concluded that SN 2012dn 
 could be classified as a SC SN. Here, we support their 
 results by focusing on the NIR observational properties.


 The light curves a normal SN Ia exhibited, the characteristic structures
 around $t=20-30$ d, e.g., the shoulder-like structure of the $R$-band.
 For normal SNe Ia, the secondary maximum magnitudes of the $I$ and $J$-band
 light curves were much fainter than the first peak. 
 On the other hand, SNe 2012dn and 2009dc exhibited much flatter shapes
 in all bands than those of normal SN 2001el. Moreover, the shapes of the 
 light curves of SNe 2012dn and 2009dc up to $t=40$ d were extremely similar. 

 Similar trends were also exhibited by the color evolutions.
 The $V-J$, $J-H$ and $J-K_{s}$ color evolutions demonstrated 
 an excellent match with those of SC SN 2009dc, while remaining quite distinct 
 from those of normal SNe Ia. Contrastingly, $V-H$ and $V-K_{s}$ color evolutions
 were well similar to those of SNe 2001el and 2009dc, while the $V$, $H$,
 and $K_{s}$-band light curves were quite different.
 The NIR light curves and color evolutions among the SNe Ia exhibited greater 
 homogeneous characteristics than those of the optical \citep{Jack2012,Dhawan2015}. 

 The estimated ejecta mass ranged from $\sim1.8-2.2~M_{\odot}$ using 
 various approaches (see Appendix 1), indicating that the total ejected mass 
 could be much larger than the Chandrasekhar-limiting mass of a 
 non-rotating WD (but, see \cite{Parrent2016}). Such a mass range for SN 2012dn
 is roughly 
 consistent with those derived from the light curve and 
 spectral analysis of SC SNe 
 \citep{Scalzo2010,Silverman2011,Taubenberger2011,Kamiya2012,Hachinger2012}.
  We conclude that SN 2012dn should be classified as a SC SNe.



 \section{Discussion}

 \subsection{NIR excess}

  SN 2012dn shows an additional and extremely long-duration 
 component in the NIR light curves.
 In the following sections, we extract this NIR component
 using light curve subtraction, and analyze the spectral
 energy distribution using dust models.

  \subsubsection{NIR light curve subtraction}

 The subtraction of the $J$, $H$ and $K_{s}$-band fluxes by those of another SC 
 SN 2009dc, was performed to derive the origin of the NIR 
 excess. 
 The $J$, $H$, and $K_{s}$-band light-curve shapes exhibited the excellent matches to 
those of SN 2009dc up to $t=40$ d, indicating that the intrinsic 
NIR light-curve behavior is almost the same to that of SN 2009dc while the absolute 
luminosities were different. From this striking similarity, we hereafter assume 
that the behavior intrinsic to the SN components are the same
between SNe 2012dn and 2009dc, and we discuss the origin of the difference 
between two SNe. While we believe this is a reasonable assumption, 
further justification of this assumption will require a larger sample of NIR 
light curves of SC SNe Ia, highlighting the importance of intensive NIR 
follow-up observations of SC SNe Ia.

 First, the $J$, $H$, and $K_{s}$-band absolute magnitudes 
 of SNe 2009dc and 2012dn were measured at their $B$-band maximum. 
 The fluxes were converted from these magnitudes using the FWHMs of the 
 filter passbands and zero magnitude fluxes \citep{Fukugita1996,Bessell1990}. 
 The fluxes of SN 2012dn were subtracted by those of SN 2009dc. 
 Stretching of the light curves
 was not performed because the $JHK_{s}$-band light curves of SN 2012dn 
 were similar to those of SN 2009dc up to $t=30$ d (see \S 4).
 When the fluxes of SN 2009dc 
 were absent during the same epoch as those of SN 2012dn, 
 the interpolation was performed for the fluxes of SN 2009dc. 
 The subtracted fluxes were converted 
 into absolute magnitudes and are plotted in Fig. 8.

 The subtracted $H$ and $K_{s}$ -band 
 light curves exhibited linear rises until $t=40$ and $50$ d, respectively,
 and the $H$ and $K_{s}$-band 
 peak magnitudes reached $-18$ and $-18.5$ mag, respectively. Thereafter, 
 they exhibited notable plateaus in the 
 $H$ and $K_{s}$ bands.  
 The subtracted $J$-band component began to be visible at a magnitude of 
 $-15.5$ after $t=90$ d,
 after the SN emission became sufficiently faint. 
 Such plateau-shaped light curves could naturally be explained 
 by the NIR echo from the CS dust \citep{Maeda2015}. 
 Comparisons of the 
 subtracted light curves with the theoretical model will be 
 performed in \S 5.2.2.

 \subsubsection{Dust model fitting}

 The SEDs of the subtracted $H$ and $K_{s}$-band 
fluxes were constructed from the effective wavelengths of the passband of their filters 
\citep{Bessell1990,Fukugita1996}. 
  Fitting to the subtracted SEDs using the dust model from \citet{Kawabata2000} 
 was then performed adopting the following procedure. 
 The dust emission SEDs were calculated using the Mie theory for dust opacities.
 The optically thin dust emissions were integrated on the assumption of 
 a distribution of the dust sizes; the power-law index was assumed to be 3.5.
 We calculated two cases of the dust composition: 
 amorphous carbon and astronomical silicate
 \citep{Kawabata2000}. 
 The dust temperatures and masses were given as variable free parameters.
 The temperatures were varied at every $100$ K in the range between $500$ and 
 $2500$ K, and the dust masses varied from 10$^{-6}$ to 10$^{-4}$ $M_{\odot}$.
 To derive the most reliable parameters, a least-squares method was adopted.
 The SED and dust model evolutions are shown in Fig. 9.
 We found that SED fitting after $t=45$ d was more reliable. 

  The temperature evolution was derived from the SED fitting analysis. 
 Figure 10 shows that the dust temperature evolution 
 exhibited an exponential decrease between $t=40$ and $100$ d. 
 In particular, it seems to exhibit a flat evolution after $t=60$ d.
 Figure 10 also shows that the dust mass evolution displayed an 
 increase from $10^{-6}$ to 10$^{-4}$ $M_{\odot}$. 
 It appears to show a flat evolution from $t=60$ d, although 
 this is later than the beginning of the $K_{s}$-band plateau phase.
 Fitting to the $J$, $H$, $K_{s}$-band fluxes was
 also performed between $t=90$ and $110$ d. 
 These results 
 may be more reliable than just the $H$ and $K_{s}$-bands because the SN 
 component was much fainter than the dust emission at this point.
 We will discuss this in the following section, 
 assuming that the fitting result is significant only after $t=45$ d.

 
 
  Until now, amorphous carbon was used in the analysis of the SED fits. 
 We also attempted SED fitting at 
 similar epochs using the astronomical silicate model \citep{Kawabata2000}. 
 Figure 9 shows a comparison of the dust emission SED of 
 amorphous carbon with that of astronomical silicate at $t=70$ d. 
 The temperatures used to explain the SEDs among these phases ranged from
 $1200-2200$ K and were significantly higher than 
 the evaporation temperature of astronomical silicate, $\sim1000$K \citep{Nozawa2003}.
 This means that the dust that accounts for the NIR excess of SN 2012dn 
 could almost completely be dominated by amorphous carbon. We will discuss 
 the origin of this dust in the following section, and adopt 
 amorphous carbon with a grain size of $0.01\mu$m.

\begin{figure*}
  \begin{center}
    \begin{tabular}{c}
      \resizebox{120mm}{!}{\includegraphics{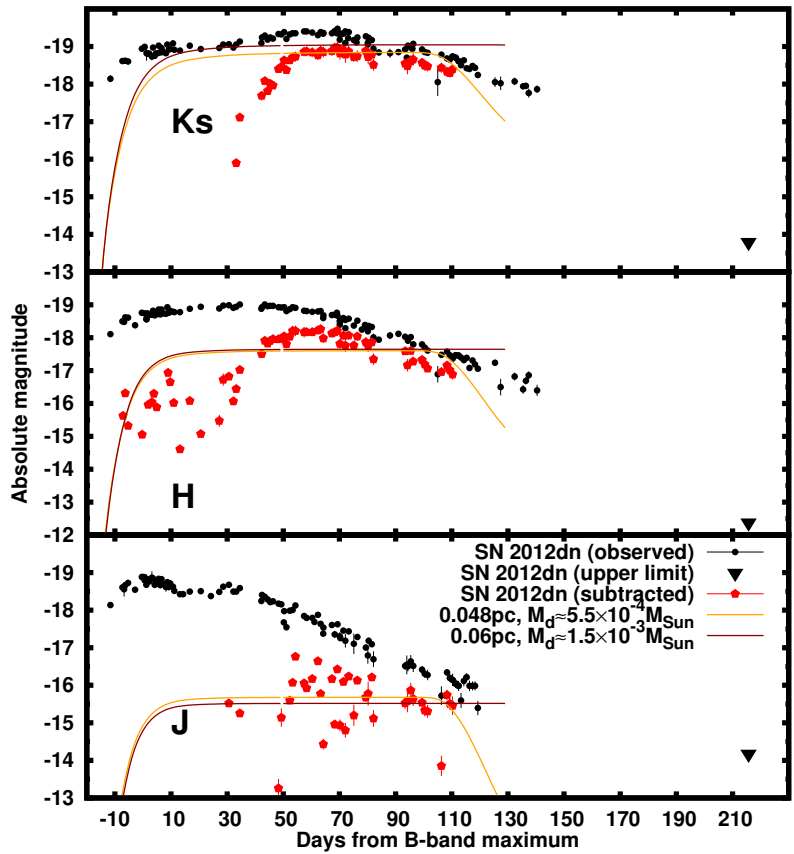}} \\
    \end{tabular}
  \end{center}
    \caption{Comparison of the absolute magnitude light curves of 
 SN 2012dn with the theoretical NIR echo models \citep{Maeda2015} using two different sets of the parameters, 
 and assuming a symmetrically spherical shell and amorphous carbon. 
 The light curves 
 subtracted by those of SN 2009dc are also plotted and are denoted by the 
 red-filled pentagon. The upper-limit magnitudes observed by IRSF telescope were also plotted 
 and denoted by the down triangles.} 
    \label{epl}
\end{figure*}

  \subsection{Origin of the dust}

  The near-infrared excess related to the dust has previously been 
 discussed for several core collapse SNe 
 \citep{Bode1980,Dwek1983a,Dwek1983b,Gerardy2000,Pozzo2004,Mattila2008,
 DiCarlo2008,Fox2009,Andrews2011,Maeda2013,Gall2014}. 
 However, there have been only a few discussions regarding
 dust signatures for the 
 subclass of SNe Ia, i.e., SN 2005gj-like events \citep{Fox2013}.
 SN Iax 2014dt also exhibited a possible mid-infrared (MIR) excess in 
 the intermediate phase, suggesting that dust emission could arise from an NIR 
 echo or other mechanisms \citep{Fox2016}. 
 They also presented a $K_{s}$-band light curve which might show a glowing excess, while 
 they did not discuss the origin of the possible $K_{s}$-band excess.
 The scenarios that could explain 
 the NIR excesses include (i) ejecta cooling 
 condensations, (ii) light echo from the interstellar/circumstellar 
 medium, and (iii) dust formation through the interaction
 of the ejecta with the CSM. For SN 2012dn, no emission lines
 were found in its early spectrum, indicating that the 
 ejecta-CSM interaction scenario is unlikely.

\subsubsection{Dust formation} 

  The NIR echo scenario could be naturally acceptable by the fact that 
 the NIR 
 light curves subtracted by those of SN 2009dc exhibited 
 definitively trapezoidal shapes (see \S 5.2.2). 
 The dust formation scenario does not 
 explain the serendipitous reproduction of such plateau evolutions. 
 Most SNe that undergo dust formation exhibit a significant increase 
 in the NIR fluxes, e.g., as observed in SN IIn 2005ip \citep{Fox2009}. 
 The blueshift of the emission lines, the sudden drop in the optical 
 luminosity, a slight increase in the NIR emission, the reddening of 
 the optical color, and the 
 evolutions of the optical depth were not found in observations of
 SN 2012dn.
 To clarify these points, we discuss in this section whether the dust formation scenario can explain the 
 NIR excess observed in SN 2012dn or not. 

  Recently, it has been reported that SC SNe have a relatively 
 fast decline in their optical light curves 
 between $t=200$ and $400$ d \citep{Maeda2009a,Taubenberger2011}.  
 Analysis suggests that this fast decline may be related to 
 dust formation scenario. 
  For SN 2012dn, \citet{Chakradhari2014} reported that the optical 
 light curves exhibited a faster evolution between $t=50$ and $100$ d 
 compared with those of other SC SNe 2007if and 2009dc, but were similar 
 to those 
 of SN 2006gz. The authors suggested that dust condensation may occur 
 from $t=50$ d, which was much earlier than that discussed for other 
 SC SNe.

  Indeed, the optical-NIR integration light curve exhibited 
 a similar 
 decline rate to that of SN 2009dc (see \S 3.3). Such serendipitous
 consistency may be explained by re-emission from the newly formed dust
 absorbing the SN emission. 
 However, it is not clear whether the consistent evolution of the 
 SN light curves are intrinsic or not. 

  From a theoretical point of view, \citet{Nozawa2011} predicted that 
 dust formation should occur 
 at a much later epoch, i.e., a year after the explosion. 
 Furthermore, a higher-density ejecta 
 relative to a normal SN Ia could further delay dust formation.
 Therefore, the timescale of the NIR increase seems to be too short 
 to be consistent with the dust formation scenario.

  If typical-sized dust forms via 
 ejecta cooling, the bluer band color should experience more substantial extinction. 
 In our case, the $U-B$ color did not exhibit 
 significant reddening between $t=30$ and $60$ d \citep{Chakradhari2014}, and
 the evolution was similar to that of SN 2009dc.
 On the other hand, the $B-V$ color was $\sim0.4$ mag redder than that 
 of SN 2009dc between $t=30$ an $60$ d. Other optical colors were 
 similar to those of SN 2009dc \citep{Chakradhari2014}. 
 Therefore, if the dust formation would be responsible to the fast 
 decline in the optical light curve, anomalous dust with 
 wavelength independent opacity is needed to 
 explain these similar color evolutions, e.g., a gray dust.

  SED fitting analysis either does not favor the dust formation 
 scenario. The optical depth evolution could show a rapid increase if 
 dust formation occurs in the ejecta, as observed in Type Ibn 
 SN 2006jc \citep{DiCarlo2008,Mattila2008,Sakon2009}.
 The optical depth did not exhibit the dramatical evolution 
 among $t=40$ and $70$ d, suggesting that 
 mass evolution by the newly formed dust was unconfirmed.
  The dust mass evolution also exhibited a flat evolution 
 until $t=100$ d, implying that no new dust formation occurred 
 between $t=40$ and $100$ d (see Figure 10). 
 All these arguments presented in this subsection support the NIR 
 echo scenario rather than the dust formation.

  \subsubsection{Light echo from circumstellar dust grains}
  

  The subtracted $H$ and $K_{s}$-band light curves exhibited 
 flat evolutions between $t=50$ and $110$ d. A NIR 
 echo could be the most plausible scenario to explain such 
 flat evolutions \citep{Dwek1983b,Chevalier1986}. 
 A comparison 
 of the subtracted light curves with those predicted by the 
 theoretical model \citep{Maeda2015} was performed in two cases. 
 Estimating the distances to the CS dust using the epoch of the 
 beginning of the NIR echo and the duration of the plateau, 
 we investigated the distance over which 
 the dust can survive in the SN radiation field. 

  The duration of a light curve is determined by the spatial extent
 of the interstellar/circumstellar dust. Dust emission 
  that is thermalized by the radiation of the central SN is seen 
  by an observer.
  For a sample of normal SNe Ia, \citet{Maeda2015} provided upper limits for 
  the CS dust masses and their distances from SNe 
  by comparing the predicted NIR echo light curves with observations.
  We attempted to compare the $JHK_{s}$-band light curves 
  of SN 2012dn with the theoretical light curves as shown in Fig. 8.
  To explain the subtracted $J$, $H$ and $K_{s}$-band luminosity,  
  the dust mass should be $5.5\times10^{-4}-1.5\times10^{-5}$~M$_{\odot}$,
  with the dust 
  assumed to have a shell-like geometry and consist of amorphous carbon. 
  The distances to the dust in the model were assumed to 
  be $\sim4.8\times10^{-2}-6.0\times10^{-2}$~pc. It should be 
  emphasized that 
  this model can simultaneously explain all NIR-band luminosity with 
  consistent masses, indicating the consistency in the 
  dust temperature between the model and in SN 2012dn.
  However, it does not explain the rising part of the subtracted light curves.

   A spherically symmetric shell cannot explain the time delay at the 
  beginning of the NIR echo. The dust would not be distributed 
  around the crosspoint between the line-of-sight direction and the circular. 
  The smallest distance was calculated to be $\sim2.0\times10^{-2}$ pc if the 
  dust was located on the opposite side of the SN to the 
  Earth, although such geometry is highly unlikely. 
  The longest distance is limited by the intensity of the SN radiation and the 
  temperature of the dust (this will be discussed in the following paragraph).
  Thus, we suggest that a reasonable distance could be 
  greater than $2.0\times10^{-2}$ pc, which is not inconsistent with the 
  distances estimated from the theoretical model. An 
  axisymmetrically circular arc shell 
  could explain such a delay. Alternatively, a highly inclined
  ring-like shell could be also considered.
  Future work on the NIR echo emission for a SC SN Ia will deal with 
  such problems. The nature of the pre-explosion system is discussed,
  using these distances and CS dust shell geometries, in \S 5.3.


\begin{figure}
  \begin{center}
    \begin{tabular}{c}
      \resizebox{80mm}{!}{\includegraphics{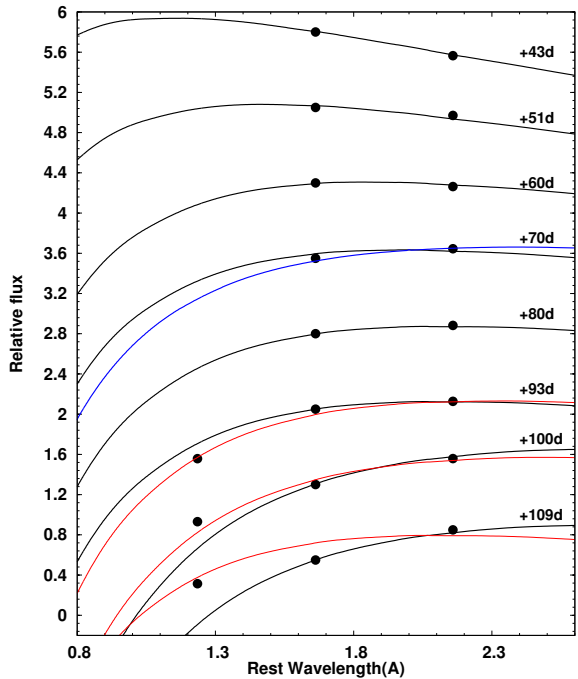}} \
    \end{tabular}
  \end{center}
    \caption{Time evolution of the spectral energy distributions (SEDs) of 
SN 2012dn between $t=43$ and $109$ d. The SEDs were converted from the 
 light curves subtracted by those of SN 2009dc.
Each epoch is denoted
at the top right side of each SED. Each line denotes a 
dust emission model \citep{Kawabata2000}. 
 The dust emissions were calculated by integrating the optically thin
 dust using an assumed power-law size distribution. The maximum size was 0.01$\mu$m.
 The dust was composed of amorphous carbon. The blue line denotes a model for
 astronomical silicate whose temperature significantly exceeds its 
 evaporation temperature. The black lines denote the fitting results for the 
 $H$ and $K_{s}$ -band fluxes, while the red line denotes the $J$, $H$, and 
 $K_{s}$-bands.} 
    \label{epl}
\end{figure}

\begin{figure}
  \begin{center}
    \begin{tabular}{c}
      \resizebox{85mm}{!}{\includegraphics{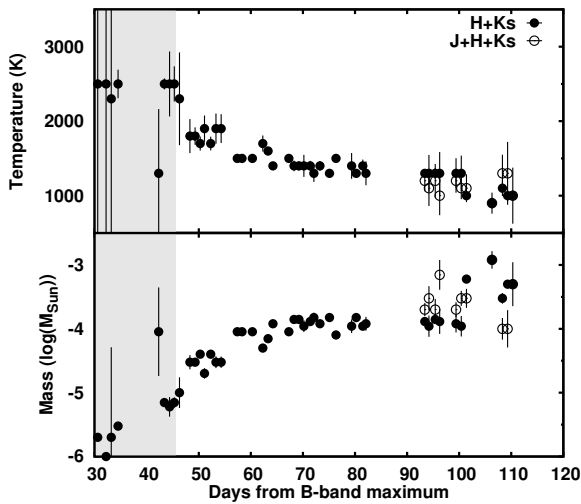}} \\
    \end{tabular}
  \end{center}
    \caption{
(Top panel) Time evolution of the dust temperatures estimated from the SED 
fitting analysis. The parameters used for the dust emission models are the same as 
those presented in Fig. 9. The estimated values in the shaded regions
are unreliable. (Bottom panel) The dust mass evolutions estimated by the same 
method as the upper panel. } 
    \label{epl}
\end{figure}

  The minimum dust cavity size can also be limited using the peak 
 luminosity because the strong radiation field can break up the 
 dust particles \citep{Dwek1983b}.  
 The scaling method for the parameters of 
 the well-studied SN to those of SN 2012dn is used along with Eq. (5) 
 presented in \citet{Fox2009}.
 SED fitting estimated the dust temperature 
 to be $T_{d}\sim1300$ K at the plateau phase. 
 The peak luminosity was already estimated to be $L=1.2 \times 10^{43}$ 
 erg~s$^{-1}$ from the $BVRIJHK_{s}$-band integration 
 light curves (see \S 3.3). 
 We used the following parameters for the scaling method:
 a dust cavity size of $r=6.0\times10^{16}$ cm, 
 temperature of $T_{d}=1900$ K, and 
 $L_{peak}=1.4\times10^{43}$ erg~s$^{-1}$ for Type II SN 1980K as a template 
 \citep{Dwek1983b}.
 The size of the dust cavity was estimated to be $\sim1.4\times10^{17}$ cm, 
 which corresponds to $4.6\times10^{-2}$ pc. 
 This is consistent with the size estimated from the theoretical NIR echo model.
 

  We have presented an illustration of the NIR echo scenario by comparing the subtracted 
 light curves with those predicted by the theoretical model.
 The distance ($\sim4.8-6.0\times10^{-2}$ pc) estimated from 
 the theoretical model 
 was consistent with the lower limit ($\sim4.6\times10^{-2}$ pc ) 
 set by the condition of the dust evaporation. Indeed, this is in support of the NIR echo scenario where the most 
 significant contribution to the NIR LC comes from the dust at the evaporation radius, 
 even if the CSM is distributed smoothly from the SN vicinity.

\subsection{Implications for the progenitor system}

   A non-rotating WD explosion in the single-degenerate scenario 
 cannot reproduce the 
 super-Chandrasekhar-limiting mass. There are two possibilities to account 
 for the large ejecta mass: a rapidly-rotating WD or a 
 double WD merger
 scenario. In a recent hydrodynamic explosion model, the extremely large 
 luminosity (i.e., the large $^{56}$Ni mass) favors 
 the detonation-triggered spherical explosion of a rapidly
 rotating WD in the single-degenerate scenario \citep{Pfannes2010}, 
 while a violent merger-induced explosion from the double-degenerate scenario 
 would explain the subluminous 
 \citep{Pakmor2010} or normal SNe Ia \citep{Pakmor2012}.

  A recurrent nova is an explosive event caused by thermonuclear runaway on the 
 surface of the WD by accretion from a companion star, e.g., a red giant star. 
 If the mass accretion from the companion is superior to the mass loss by the 
 eruption, the WD mass grew up to the limiting mass. Thereafter, the thermonuclear
 explosion of the entire WD material should occur \citep{Hachisu1999,Hachisu2012}.
 The Roche lobe overflow 
 caused by the binary interaction also leads to a dense environment 
 surrounding the system. 
 The single-degenerate scenario favors such a dense environment around the 
 progenitor. 
 
  The envelope 
 predicted by the merger scenario would be expected to be ejected within $\sim1$ day
 after the disruption of the secondary star 
 \citep{Yoon2007,Levanon2015,Tanikawa2015}. The interaction of the ejecta 
 with such a close envelope may rather predict the UV excess at the earlier 
 phase \citep{Scalzo2010,Scalzo2012,Brown2014a,Brown2014b}. 
 The density on the subparsec scale is low because the time scale
 over which the merger event of the two WDs occurs is very long 
 due to the formation of the WD binary.

  Our analysis found that the NIR echo light curve of SN 2012dn between 
 $t=40$ and $110$ d is well explained by the existence of CS 
 material surrounding the pre-explosion system in the single-degenerate scenario.
 Such an NIR echo implies that the pre-explosion material existed at a distance
 greater than $\sim4.6\times10^{-2}$ pc.
 Such distant material strongly supports the presence of a wind from 
 the pre-explosion system, 
 e.g., a recurrent nova. 

  The mass loss rate of the pre-explosion system can be estimated 
 within the framework of 
 the NIR echo scenario. The radius and 
 the total mass of the CS dust are required. The former 
 was calculated using the timescale over which the NIR echo luminosity 
 sufficiently decreased, and was adopted to be 
 $R=1.8\times10^{17}$ cm. 
 The total mass was adopted as $M_{d}=\sim5.5\times10^{-4}-1.5\times10^{-3} 
 M_{\odot}$
 from the comparison with the theoretical NIR echo. 
 Given that the model overpredict the early rising light curve,
 the mass should be reduced by a factor $\sim2$.
 The gas-dust mass ratio was assumed to be $Z_{d}=0.01$. 
 The mass loss
 rate was calculated to be $4.8\times10^{-6}-1.3\times10^{-6}M_{\odot}$ 
 yr$^{-1}$, 
 assuming that the shell velocity was 10 km~s$^{-1}$. 
 The estimated mass loss rate ($\sim10^{-6}$-$10^{-5}M_{\odot}$ yr$^{-1}$) 
 is roughly consistent with those measured from Galactic 
 recurrent novae \citep{Hachisu2000,Hachisu2001}.

  The deep images of Galactic SN Ia progenitor candidates 
 V445 pup and RS Oph obtained several years after 
 their eruptions have been reported \citep{Woudt2009,Bode2007}. 
 Axisymmetric bipolar shells formed by the ionized gases 
 were found, and these exhibited a mildly spread opening angle.
 The time delay and durations of the NIR echo in SN 2012dn
 are likely to be explained by such geometry even if the line-of-sight
 direction is parallel to the CS shell plane.
 Another NIR echo from the opposite side of the shell surrounding 
 SN 2012dn would reach us in the near future, depending on the geometry. 
 Alternatively, 
 the interaction of emission from the collision of the ejecta 
 with the shell may be discovered. Assuming that the ejecta 
 velocity is $\sim10,000$~km~s$^{-1}$, the interaction will occur 
 in the early 2017.






\section{Conclusion}

  In this report, we present the $JHK_{s}$-band light curves of SN 2012dn
 exhibited 
 maxima with unusually 
 long durations. The color exhibited a strong NIR excess from 
 30 days after the $B$-band maximum. None of these phenomena have previously 
 been detected among SC SNe Ia so far, nor normal SNe Ia.

 NIR light curves subtracted by those of SN 2009dc displayed flat 
 evolutions, indicating a NIR echo from the surrounding dust shell to 
 the SN. The fitting of the SEDs was performed using a dust emission model with 
 amorphous carbon and a maximum grain size of $0.01{\mu}$m. The dust mass
 evolution also exhibited a plateau phase between $t=60$ and $110$ d.
 Comparison of the subtracted light curve with the theoretical model 
 provided a dust mass of $5.5\times10^{-4}-1.5\times10^{-3} M_{\odot}$ 
 and a distance 
 of $4.8-6.0\times10^{-2}$ pc, suggesting the presence of circumstellar 
 material in the pre-explosion system. 

  If the pre-explosion system had 
 a wind of velocity of 10-100 km~s$^{-1}$, the mass loss rate is 
 estimated to be $10^{-6}-10^{-5}M_{\odot}$ yr$^{-1}$, which is
 well consistent with those measured in Galactic recurrent 
 novae. 
 The light curve and color evolution comparisons indicate that 
 SN 2012dn should belong to the class of SC SNe, while the total ejecta mass could be
 1.8-2.2$M_{\odot}$.  
 We conclude that a SC SN should originate from a WD explosion in 
 the single-degenerate scenario.

\begin{ack}
  This work was supported
 by the Optical and Near-infrared Astronomy Inter-University Cooperation Program, 
 and the Hirao Taro Foundation of the Konan University Association 
 for Academic Research. This work was partly supported by the Grant-in-Aid for 
 Scientific Research from JSPS (26800100) and MEXT (24103003). 
 The works by K.M., N.T., and M.T. are partly supported by World Premier International 
 Research Center Initiative (WPI Initiative), MEXT, Japan. 
\end{ack}

\appendix
 \section{Estimate of the ejecta mass}
  The total ejecta mass was estimated with a scaling method using 
 Arnett's rule \citep{Arnett1982} by comparison with a normal SN Ia.
 The slower light curve evolution indicates a greater 
 ejecta mass if the line velocities are similar. The line velocity of
 SN 2012dn was almost the same as that of a low-velocity SN Ia at a 
 similar epoch while
 the light curve evolutions were much slower. This suggests that the total 
 ejecta mass could be significantly larger than that of a normal SN Ia.

 \citet{Tanaka2011} presented an ejecta mass of $\sim1.4M_{\odot}$ and 
 kinetic energy $\sim1.3\times10^{51}$ erg~s$^{-1}$ for normal SN 2003du. 
 These ejecta properties were adopted as zero points in the scaling method
 The velocities obtained by the same absorption lines were used for the 
 scaling. We adopt the S~{\sc ii} velocities, since the S~{\sc ii} 
 line formation region
 was located in the inner part of the ejecta compared with that of Si{\sc ii}
 (see Fig. 7). 
 The line velocities of these SNe were derived as  
 $v_{\rm S~{\sc ii},12dn}\sim9500$ km~s$^{-1}$ and 
 $v_{\rm S~{\sc ii},03du}\sim9000$ km~s$^{-1}$ at the $t=5$ d. 
 Arnett's rule is defined as, 
 \begin{equation}
 \tau \propto M_{ej}^{3/4} \cdot E_{K}^{-1/4}, 
 \end{equation}
 \begin{equation}
 v \propto M_{ej}^{-1/2} \cdot E_{K}^{1/2}.
 \end{equation} 
 Here, $\tau$ is the timescale of the light curve and $v$ is the 
 expansion velocity.  
 The light curve timescale was adopted to describe the decline rate of the 
 $B$-band light curves. The ratio was given as 
 $\tau_{12dn}\sim1.08$ and $\tau_{03du}\sim0.83$ from the inverse 
 of the $\Delta$~$m_{15}$($B$)$_{12dn}=0.92$ and $\Delta$~$m_{15}$($B$)$_{03du}=1.02$ 
 (corresponding to the stretch factors). 
 The total ejecta mass was estimated to be $M_{ej,12dn}\sim1.8M_{\odot}$
 from $M_{ej,03du}=1.4M_{\odot}$. The total ejected mass 
 significantly exceeds the Chandrasekhar-limiting mass of the WD. 

  A cross check was performed using scaling from the parameters
 of another SC SN 2009dc. An ejecta mass of $\sim2.4M_{\odot}$ and 
 $E_{k}=0.9\times10^{51}$ erg~s$^{-1}$ were adopted as zero points 
 \citep{Kamiya2012}. 
 The line velocity of S~{\sc ii} at $t=5$ d was measured to be 
 $\sim6000$ km~s$^{-1}$ \citep{Yamanaka2009a}
 and the decline rate was $\Delta$~$m_{15}$($B$)$\sim0.7$ 
 \citep{Yamanaka2009a,Silverman2011,Taubenberger2011}.  
 The total ejecta mass of SN 2012dn was estimated to be 
 $\sim2.2M_{\odot}$, which is slightly larger than that estimated above.

\begin{table*}
\scriptsize
    \caption{Summary of the properties of the telescopes, instruments, and observatories.} 
\begin{center}
\begin{tabular}{lllll}
    \hline 
    \hline 
     Observatory     &  Telescope           &  Instruments & Filters/Resolutions                  &  Number of nights \\
    \hline 
     NO$^{a}$        &  1.6m Pirka          &  MSI$^{b}$      & $U$, $B$, $V$, $R$, $I$           & 44    \\
                     &                      &                 & $R=150$                           & 3     \\
     AO$^{c}$        &  0.5m MITSuME$^{d}$  &  CCD            & $g'$, $R$, $I$                    & 11    \\
     KAO$^{e}$       &  1.3m Araki          &  Adler$^{f}$    & $B$, $V$, $I$, $g'$, $i'$, $z'$,  & 24    \\
     OKUAO$^{g}$     & 0.5m                 &  Andor CCD      & $B$, $V$, $R$, $I$                & 15    \\
                     &                      &  ST10 CCD       & $B$, $V$, $R$, $I$                & 14    \\
     OAO$^{h}$       & 0.5m MITSuME$^{d}$   &  CCD            & $g'$, $R$, $I$                    & 67    \\
                     & 188cm                &  KOOLS$^{i}$    & $B$, $V$, $R$, $I$                & 1     \\ 
     NHAO$^{j}$      & 2.0m Nayuta          &  NIC$^{k}$      & $J$, $H$, $K_{s}$                 & 21     \\
     HHO$^{l}$       & 1.5m Kanata          &  HOWPol$^{m}$   & $B$, $V$, $R$, $I$, $z'+Y$        & 52    \\
                     &                      &                 & $R=400$                           & 30    \\
     IO$^{n}$        & 1.0m                 &  IR Cam         & $J$, $H$, $K_{s}$                 & 24    \\
     IAO$^{o}$       & 1.05m MITSuME$^{d}$   & CCD             & $g'$, $R$, $I$                    & 29     \\
     SAAO$^{p}$      & 1.4m IRSF            & SIRIUS$^{q}$    & $J$, $H$, $K_{s}$                 & 80    \\
    \hline
\end{tabular}{}
\end{center}
{\footnotesize
{\bf Note.} 
$^{a}$Nayoro Observatory;
$^{b}$Multispectral Imager \citep{Watanabe2012};
$^{c}$Akeno Observatory;
$^{d}$Multicolor Imaging Telescopes for Survey and Monstrous 
 Explosions \citep{Kotani2005};
$^{e}$Koyama Astronomical Obsrvatory;
$^{f}$Araki telescope DuaL-band imagER \citep{Isogai2015};
$^{g}$Osaka Kyoiku University Astronomical Observatory;
$^{h}$Okayama Astrophysical Observatory;
$^{i}$Kyoto Okayama Optical Low-dispersion Spectrograph \citep{Ohtani1998};
$^{j}$Nishi-Harima Astronomical Observatory;
$^{k}$Nishi-harima Infrared Camera;
$^{l}$Higashi-Hiroshima Observatory;
$^{m}$Hiroshima One-shot Wide-field Polarimeter \citep{Kawabata2008};
$^{n}$Iriki Observatory;
$^{o}$Ishigaki-jima Astronomical Observatory;
$^{p}$South African Astronomical Observatory;
$^{q}$Near-infrared simultaneous three-band camera \citep{Nagayama2003}
}
\end{table*} 

\begin{center}
\scriptsize
\begin{longtable}{lllllll}

    \caption{MSI photometry of SN 2012dn in the $UBVRI$ bands} \\
\hline
\hline
MJD     & Phase & $U$         & $B$         & $V$     & $R$ & $I$ \\
\hline
\endhead
56122.6 & -9.79 & 14.513(0.098) & 15.172(0.016) & 15.005(0.013) & 14.974(0.025) & 14.879(0.025) \\
56123.6 & -8.79 & 14.379(0.098) & 14.990(0.016) & 14.856(0.011) & 14.841(0.025) & 14.759(0.026) \\
56124.6 & -7.79 & 14.262(0.100) & 14.881(0.017) & 14.747(0.013) & 14.726(0.035) & 14.666(0.026) \\
56125.6 & -6.79 & 14.169(0.101) & 14.797(0.017) & 14.640(0.012) & 14.639(0.025) & 14.596(0.031) \\
56126.6 & -5.79 & 14.142(0.099) & 14.597(0.020) & 14.438(0.017) & $\cdots$ & $\cdots$ \\
56128.6 & -3.79 & 14.009(0.099) & 14.551(0.020) & 14.448(0.012) & 14.442(0.025) & 14.429(0.025) \\
56129.6 & -2.79 & $\cdots$ & $\cdots$ & 14.373(0.025) & $\cdots$ & $\cdots$ \\
56131.6 & -0.79 & 13.999(0.099) & 14.532(0.017) & 14.360(0.012) & 14.334(0.025) & 14.360(0.036) \\
56136.6 & 4.21 & 14.298(0.102) & 14.543(0.017) & 14.309(0.012) & $\cdots$ & $\cdots$ \\
56137.7 & 5.31 & $\cdots$ & $\cdots$ & 14.344(0.019) & $\cdots$ & $\cdots$ \\
56138.6 & 6.21 & 14.321(0.110) & 14.630(0.017) & 14.373(0.013) & 14.357(0.026) & 14.373(0.040) \\
56142.6 & 10.21 & $\cdots$ & $\cdots$ & 14.456(0.049) & $\cdots$ & $\cdots$ \\
56145.6 & 13.21 & $\cdots$ & $\cdots$ & 14.501(0.013) & $\cdots$ & $\cdots$ \\
56146.5 & 14.11 & 15.069(0.098) & 15.249(0.017) & 14.542(0.012) & $\cdots$ & $\cdots$ \\
56147.6 & 15.21 & 15.203(0.099) & 15.326(0.017) & 14.603(0.012) & 14.504(0.025) & 14.320(0.027) \\
56148.6 & 16.21 & 15.339(0.099) & 15.487(0.017) & 14.694(0.018) & 14.575(0.025) & 14.370(0.026) \\
56150.5 & 18.11 & 15.635(0.100) & 15.677(0.017) & 14.778(0.016) & 14.631(0.026) & 14.378(0.026) \\
56153.5 & 21.11 & 16.031(0.109) & 16.009(0.020) & 14.958(0.013) & 14.701(0.025) & 14.357(0.026) \\
56155.5 & 23.11 & 16.302(0.100) & 16.226(0.016) & 15.110(0.011) & 14.784(0.025) & 14.377(0.025) \\
56157.6 & 25.21 & 16.298(0.141) & 16.412(0.018) & 15.145(0.012) & $\cdots$ & $\cdots$ \\
56160.5 & 28.11 & 16.622(0.195) & 16.639(0.016) & 15.299(0.012) & $\cdots$ & $\cdots$ \\
56161.5 & 29.11 & 16.873(0.100) & 16.719(0.018) & 15.350(0.012) & $\cdots$ & $\cdots$ \\
56166.5 & 34.11 & 16.950(0.120) & 16.957(0.020) & 15.552(0.023) & $\cdots$ & $\cdots$ \\
56167.5 & 35.11 & $\cdots$ & 17.102(0.104) & $\cdots$ & $\cdots$ & $\cdots$ \\
56169.5 & 37.11 & $\cdots$ & 16.959(0.155) & 15.574(0.024) & $\cdots$ & $\cdots$ \\
56171.6 & 39.21 & $\cdots$ & 17.114(0.043) & 15.713(0.016) & $\cdots$ & $\cdots$ \\
56172.6 & 40.21 & 16.874(0.186) & 17.136(0.023) & 15.767(0.015) & $\cdots$ & $\cdots$ \\
56173.5 & 41.11 & $\cdots$ & 17.117(0.021) & 15.767(0.013) & $\cdots$ & $\cdots$ \\
56174.5 & 42.11 & $\cdots$ & 17.265(0.091) & 15.823(0.013) & $\cdots$ & $\cdots$ \\
56175.5 & 43.11 & 17.431(0.123) & 17.256(0.017) & 15.856(0.013) & $\cdots$ & $\cdots$ \\
56177.5 & 45.11 & 16.618(0.293) & 17.253(0.017) & 16.037(0.013) & 15.609(0.030) & 15.094(0.029) \\
56182.6 & 50.21 & $\cdots$ & 17.309(0.033) & 16.028(0.014) & $\cdots$ & $\cdots$ \\
56184.5 & 52.11 & 17.519(0.131) & 17.474(0.027) & 16.104(0.013) & $\cdots$ & $\cdots$ \\
56187.5 & 55.11 & $\cdots$ & $\cdots$ & 16.074(0.039) & $\cdots$ & $\cdots$ \\
56188.5 & 56.11 & 17.861(0.274) & 17.612(0.022) & 16.207(0.019) & $\cdots$ & $\cdots$ \\
56189.5 & 57.11 & $\cdots$ & 17.043(0.275) & 16.191(0.025) & $\cdots$ & $\cdots$ \\
56190.5 & 58.11 & 17.887(0.124) & 17.484(0.021) & 16.412(0.014) & 16.073(0.027) & 15.567(0.025) \\
56191.6 & 59.21 & $\cdots$ & $\cdots$ & 16.270(0.052) & $\cdots$ & $\cdots$ \\
56202.4 & 70.01 & $\cdots$ & 17.852(0.042) & 16.851(0.019) & 16.574(0.029) & 16.107(0.034) \\
56205.5 & 73.11 & $\cdots$ & $\cdots$ & 16.667(0.019) & $\cdots$ & $\cdots$ \\
56207.5 & 75.11 & $\cdots$ & $\cdots$ & 16.534(0.016) & $\cdots$ & $\cdots$ \\
56224.4 & 92.01 & $\cdots$ & $\cdots$ & 17.486(0.050) & 17.252(0.052) & 16.691(0.067) \\
56226.4 & 94.01 & $\cdots$ & $\cdots$ & 17.360(0.044) & $\cdots$ & $\cdots$ \\
\hline
\end{longtable}{}
\end{center}

\begin{center}
\scriptsize
\begin{longtable}{llllll}
    \caption{MITSuME telescope photometry of SN 2012dn in the $g'RI$ bands} \\
\hline
\hline
MJD     & Phase & $g'$         & $R$         & $I$     & Observatories \\
\hline
\endhead
56120.7 & -11.7 & 15.407(0.107) & 15.366(0.088) & $\cdots$ & IAO \\
56121.7 & -10.7 & 15.162(0.107) & 15.200(0.088) & $\cdots$ & IAO \\
56123.6 & -8.8 & 14.821(0.030) & 14.806(0.031) & 14.723(0.038) & OAO \\
56123.6 & -8.8 & 14.830(0.031) & 14.834(0.033) & 14.801(0.040) & OAO \\
56123.7 & -8.7 & 14.838(0.107) & 14.883(0.088) & $\cdots$ & IAO \\
56124.6 & -7.8 & 14.544(0.025) & $\cdots$ & $\cdots$ & AO \\
56124.7 & -7.7 & 14.606(0.040) & 14.748(0.037) & 14.664(0.044) & OAO \\
56125.6 & -6.8 & 14.661(0.029) & 14.638(0.030) & 14.570(0.033) & OAO \\
56125.6 & -6.8 & 14.730(0.028) & 14.616(0.030) & 14.558(0.033) & OAO \\
56125.6 & -6.8 & 14.636(0.027) & 14.613(0.029) & 14.513(0.033) & OAO \\
56126.6 & -5.8 & 14.457(0.107) & 14.540(0.088) & $\cdots$ & IAO \\
56126.6 & -5.8 & 14.527(0.030) & 14.531(0.032) & 14.501(0.039) & OAO \\
56126.6 & -5.8 & 14.540(0.031) & 14.539(0.032) & 14.499(0.039) & OAO \\
56126.6 & -5.8 & 14.544(0.031) & 14.519(0.033) & 14.505(0.041) & OAO \\
56127.7 & -4.8 & 14.524(0.046) & 14.439(0.037) & 14.444(0.045) & OAO \\
56127.7 & -4.7 & 14.429(0.106) & 14.511(0.087) & $\cdots$ & IAO \\
56129.7 & -2.7 & 14.348(0.106) & 14.412(0.087) & $\cdots$ & IAO \\
56131.6 & -0.8 & 14.431(0.031) & 14.334(0.030) & 14.320(0.034) & OAO \\
56131.7 & -0.7 & 14.404(0.032) & 14.280(0.034) & 14.320(0.040) & AO \\
56132.6 & 0.2  & 14.363(0.028) & 14.320(0.029) & 14.346(0.033) & OAO \\
56132.6 & 0.2  & 14.393(0.027) & 14.316(0.028) & 14.325(0.032) & OAO \\
56132.6 & 0.2  & 14.389(0.027) & 14.304(0.028) & 14.326(0.032) & OAO \\
56133.6 & 1.2  & 14.391(0.030) & 14.285(0.029) & 14.307(0.033) & OAO \\
56133.6 & 1.2  & 14.367(0.030) & 14.304(0.028) & 14.285(0.033) & OAO \\
56133.6 & 1.2  & 14.380(0.027) & 14.303(0.028) & 14.320(0.032) & OAO \\
56134.6 & 2.2  & 14.251(0.107) & 14.284(0.088) & 14.285(0.020) & IAO \\
56134.6 & 2.2  & 14.341(0.026) & 14.278(0.028) & 14.308(0.032) & OAO \\
56134.6 & 2.2  & 14.384(0.026) & 14.293(0.028) & 14.295(0.032) & OAO \\
56134.6 & 2.2  & 14.364(0.026) & 14.287(0.028) & 14.304(0.032) & OAO \\
56135.6 & 3.2  & 14.374(0.038) & 14.312(0.030) & 14.320(0.036) & OAO \\
56135.7 & 3.3  & 14.329(0.106) & 14.332(0.087) & 14.318(0.014) & IAO \\
56136.6 & 4.2  & 14.332(0.106) & 14.325(0.087) & 14.331(0.014) & IAO \\
56136.6 & 4.2  & 14.334(0.035) & 14.298(0.030) & 14.291(0.036) & OAO \\
56136.7 & 4.3  & 14.390(0.028) & 14.286(0.029) & 14.323(0.034) & OAO \\
56137.7 & 5.3  & 14.489(0.041) & 14.354(0.033) & 14.337(0.039) & OAO \\
56138.6 & 6.2  & 14.364(0.031) & 14.325(0.031) & 14.341(0.035) & OAO \\
56138.6 & 6.2  & 14.493(0.032) & 14.272(0.030) & 14.343(0.034) & OAO \\
56138.6 & 6.2  & 14.425(0.038) & 14.340(0.030) & 14.372(0.036) & OAO \\
56139.5 & 7.1  & 14.762(0.077) & 14.377(0.057) & 14.453(0.079) & OAO \\
56139.6 & 7.2  & 14.415(0.071) & 14.262(0.032) & 14.181(0.033) & AO \\
56141.6 & 9.2  & $\cdots$ & 14.286(0.064) & 14.187(0.033) & AO \\
56141.7 & 9.3  & 14.578(0.035) & 14.383(0.031) & 14.328(0.036) & OAO \\
56142.6 & 10.2 & 14.500(0.031) & 14.383(0.030) & 14.378(0.034) & OAO \\
56142.7 & 10.3 & 14.560(0.030) & 14.409(0.028) & 14.336(0.034) & OAO \\
56142.7 & 10.3 & 14.602(0.031) & 14.343(0.029) & 14.358(0.032) & OAO \\
56143.6 & 11.2 & 14.644(0.026) & 14.393(0.028) & 14.361(0.031) & OAO \\
56143.6 & 11.2 & 14.651(0.027) & 14.390(0.028) & 14.329(0.031) & OAO \\
56143.6 & 11.2 & 14.648(0.026) & 14.394(0.028) & 14.327(0.030) & OAO \\
56145.6 & 13.2 & 14.749(0.027) & 14.437(0.030) & 14.348(0.035) & AO \\
56145.6 & 13.2 & 14.755(0.028) & 14.442(0.029) & 14.434(0.032) & OAO \\
56145.6 & 13.2 & 14.757(0.028) & 14.439(0.028) & 14.392(0.032) & OAO \\
56146.6 & 14.2 & 14.878(0.038) & 14.397(0.036) & 14.252(0.046) & AO \\
56147.6 & 15.2 & 14.898(0.035) & 14.542(0.031) & 14.414(0.036) & OAO \\
56147.6 & 15.2 & 14.932(0.034) & 14.538(0.030) & 14.394(0.034) & OAO \\
56147.7 & 15.3 & 14.954(0.106) & 14.566(0.087) & 14.346(0.016) & IAO \\
56148.6 & 16.2 & 14.985(0.038) & 14.559(0.031) & 14.370(0.036) & OAO \\
56148.6 & 16.2 & 15.013(0.043) & 14.571(0.033) & 14.392(0.037) & OAO \\
56148.6 & 16.2 & 14.961(0.041) & 14.548(0.032) & 14.366(0.038) & OAO \\
56150.7 & 18.3 & 15.169(0.038) & 14.594(0.031) & 14.391(0.036) & OAO \\
56151.6 & 19.2 & 15.288(0.106) & 14.662(0.087) & 14.364(0.014) & IAO \\
56152.7 & 20.3 & 15.432(0.106) & 14.721(0.087) & 14.353(0.012) & IAO \\
56153.5 & 21.1 & 15.496(0.106) & 14.749(0.088) & 14.333(0.015) & IAO \\
56154.6 & 22.2 & 15.607(0.031) & 14.737(0.029) & 14.379(0.032) & OAO \\
56154.6 & 22.2 & 15.540(0.032) & 14.718(0.028) & 14.400(0.032) & OAO \\
56154.6 & 22.2 & 15.557(0.031) & 14.747(0.029) & 14.389(0.034) & OAO \\
56155.6 & 23.2 & 15.633(0.030) & 14.789(0.029) & 14.388(0.032) & OAO \\
56155.6 & 23.2 & 15.681(0.031) & 14.755(0.028) & 14.393(0.031) & OAO \\
56155.6 & 23.2 & 15.692(0.030) & 14.778(0.028) & 14.389(0.031) & OAO \\
56156.6 & 24.2 & 15.729(0.036) & 14.793(0.029) & 14.409(0.032) & OAO \\
56156.6 & 24.2 & 15.799(0.038) & 14.816(0.029) & 14.437(0.033) & OAO \\
56156.6 & 24.2 & 15.726(0.040) & 14.828(0.030) & 14.414(0.032) & OAO \\
56158.6 & 26.2 & 15.931(0.033) & 14.914(0.028) & 14.448(0.031) & OAO \\
56158.6 & 26.2 & 15.955(0.107) & 14.848(0.088) & 14.416(0.014) & IAO \\
56158.6 & 26.2 & 15.901(0.031) & 14.916(0.029) & 14.435(0.032) & OAO \\
56158.6 & 26.2 & 15.920(0.031) & 14.867(0.028) & 14.448(0.031) & OAO \\
56159.6 & 27.2 & 15.995(0.033) & 14.930(0.029) & 14.478(0.032) & OAO \\
56159.6 & 27.2 & 16.075(0.106) & 14.962(0.087) & 14.473(0.012) & IAO \\
56159.6 & 27.2 & 15.983(0.031) & 14.946(0.029) & 14.491(0.032) & OAO \\
56159.6 & 27.2 & 16.043(0.034) & 14.927(0.029) & 14.484(0.032) & OAO \\
56160.6 & 28.2 & 16.061(0.039) & 14.992(0.030) & 14.537(0.033) & OAO \\
56160.6 & 28.2 & 16.064(0.035) & 14.980(0.030) & 14.523(0.034) & OAO \\
56160.6 & 28.2 & 16.099(0.042) & 14.997(0.030) & 14.526(0.034) & OAO \\
56164.7 & 32.3 & 16.245(0.039) & 15.138(0.030) & 14.620(0.032) & OAO \\
56165.6 & 33.2 & 16.238(0.036) & 15.153(0.030) & 14.662(0.033) & OAO \\
56169.6 & 37.2 & 16.682(0.127) & 15.217(0.035) & 14.803(0.040) & OAO \\
56171.5 & 39.1 & 16.632(0.106) & 15.460(0.087) & $\cdots$ & IAO \\
56172.6 & 40.2 & 16.626(0.106) & 15.453(0.087) & $\cdots$ & IAO \\
56173.5 & 41.1 & 16.197(0.108) & 15.351(0.064) & 14.686(0.066) & AO \\
56174.6 & 42.2 & 16.576(0.107) & 15.470(0.087) & 14.844(0.014) & IAO \\
56178.6 & 46.2 & 16.694(0.057) & 15.643(0.034) & 15.154(0.041) & OAO \\
56178.6 & 46.2 & 16.856(0.107) & 15.759(0.087) & 15.146(0.013) & IAO \\
56179.5 & 47.1 & 16.884(0.107) & 15.800(0.088) & 15.191(0.014) & IAO \\
56180.5 & 48.1 & 16.512(0.069) & $\cdots$ & $\cdots$ & AO \\
56182.5 & 50.1 & 16.943(0.107) & 15.901(0.088) & 15.293(0.014) & IAO \\
56182.5 & 50.1 & 16.659(0.093) & 15.857(0.046) & 15.225(0.044) & OAO \\
56183.5 & 51.1 & 16.943(0.107) & 15.947(0.087) & 15.309(0.014) & IAO \\
56184.5 & 52.1 & 16.966(0.107) & 15.947(0.087) & 15.306(0.014) & IAO \\
56187.6 & 55.2 & 17.054(0.107) & 16.053(0.087) & 15.521(0.014) & IAO \\
56191.5 & 59.1 & 17.210(0.107) & 16.280(0.088) & 15.639(0.014) & IAO \\
56224.5 & 92.1 & 17.925(0.111) & 17.611(0.092) & 17.087(0.031) & IAO \\
56226.5 & 94.1 & 18.012(0.109) & 17.721(0.090) & 17.160(0.026) & IAO \\
56238.5 & 106.1 & 18.129(0.113) & 17.962(0.097) & 17.587(0.056) & IAO \\
56239.5 & 107.1 & 18.235(0.115) & 18.090(0.096) & 17.745(0.054) & IAO \\
\hline
\end{longtable}{}
\end{center}

\begin{center}
\scriptsize
\begin{longtable}{llllllll}
    \caption{Araki telescope photometry of SN 2012dn in the $g'i'z'BVI$ bands} \\
\hline
\hline
MJD     & Phase &$B$         & $V$             & $I$   &$g'$       & $i'$ &   $z'$      \\
\hline
\endhead
56125.7 & -6.7 & 14.598(0.019) & 14.544(0.011) & 14.513(0.012) & 14.523(0.014) & 14.829(0.011) & 14.988(0.000) \\
56126.7 & -5.7 & 14.511(0.019) & 14.454(0.015) & 14.454(0.016) & 14.445(0.015) & 14.790(0.012) & 14.931(0.000) \\
56131.7 & -0.7 & 14.341(0.020) & 14.269(0.014) & 14.282(0.012) & 14.248(0.015) & 14.621(0.011) & 14.650(0.000) \\
56132.7 & 0.3 & 14.355(0.032) & 14.346(0.036) & 14.260(0.041) & 14.240(0.017) & 14.607(0.012) & 14.622(0.001) \\
56133.6 & 1.2 & 14.393(0.052) & 14.201(0.025) & 14.258(0.033) & 14.251(0.015) & 14.618(0.012) & 14.478(0.003) \\
56134.7 & 2.3 & 14.359(0.021) & 14.254(0.013) & 14.264(0.012) & 14.260(0.015) & 14.613(0.013) & 14.582(0.000) \\
56135.7 & 3.3 & 14.374(0.021) & 14.215(0.016) & 14.255(0.012) & 14.254(0.014) & 14.603(0.010) & 14.568(0.000) \\
56138.6 & 6.2 & 14.461(0.019) & 14.275(0.012) & 14.265(0.012) & 14.305(0.014) & 14.618(0.010) & 14.511(0.000) \\
56139.6 & 7.2 & 14.519(0.019) & 14.286(0.013) & 14.252(0.018) & 14.335(0.016) & 14.606(0.014) & 14.481(0.000) \\
56140.7 & 8.3 & 14.577(0.019) & 14.313(0.012) & 14.263(0.016) & 14.371(0.015) & 14.625(0.013) & 14.456(0.001) \\
56141.6 & 9.2 & 14.668(0.036) & 14.350(0.015) & 14.227(0.034) & 14.424(0.018) & 14.630(0.017) & 14.444(0.003) \\
56142.6 & 10.2 & 14.730(0.020) & 14.368(0.012) & 14.276(0.014) & 14.466(0.015) & 14.656(0.012) & 14.461(0.000) \\
56143.6 & 11.2 & 14.797(0.019) & 14.380(0.012) & 14.296(0.012) & 14.505(0.015) & 14.645(0.012) & 14.439(0.000) \\
56145.6 & 13.2 & 14.983(0.020) & 14.449(0.016) & 14.304(0.013) & 14.647(0.015) & 14.682(0.012) & 14.420(0.000) \\
56146.6 & 14.2 & 15.088(0.019) & 14.492(0.017) & 14.320(0.012) & 14.722(0.015) & 14.691(0.015) & 14.420(0.000) \\
56147.6 & 15.2 & 15.195(0.020) & 14.549(0.012) & 14.328(0.011) & 14.792(0.015) & 14.704(0.012) & 14.433(0.000) \\
56148.6 & 16.2 & 15.299(0.020) & 14.598(0.011) & 14.320(0.013) & 14.877(0.015) & 14.713(0.012) & 14.426(0.000) \\
56163.6 & 31.2 & 16.584(0.024) & 15.436(0.012) & 14.535(0.011) & 16.081(0.018) & 14.934(0.011) & 14.660(0.000) \\
56174.6 & 42.2 & $\cdots$ & 15.903(0.025) & 14.923(0.013) & 16.590(0.039) & 15.344(0.015) & 15.025(0.000) \\
56181.5 & 49.1 & 17.200(0.031) & 16.084(0.017) & 15.191(0.012) & 16.747(0.018) & 15.648(0.012) & 15.291(0.000) \\
56183.5 & 51.1 & $\cdots$ & 16.076(0.040) & 15.234(0.019) & 16.794(0.036) & 15.717(0.013) & 15.376(0.000) \\
56196.5 & 64.1 & $\cdots$ & 16.593(0.018) & 15.805(0.016) & 17.095(0.022) & 16.280(0.015) & 15.976(0.000) \\
56197.5 & 65.1 & $\cdots$ & $\cdots$ & $\cdots$ & $\cdots$ & $\cdots$ & 15.947(0.000) \\
56198.5 & 66.1 & $\cdots$ & $\cdots$ & $\cdots$ & $\cdots$ & $\cdots$ & 16.075(0.001) \\
\hline
\end{longtable}{}
\end{center}

\begin{center}
\scriptsize
\begin{longtable}{lllllll}
    \caption{OKU 51cm telescope photometry of SN 2012dn in the $BVRI$ bands} \\
\hline
\hline
MJD     & Phase & $B$         & $V$             &$R$         & $I$     & Instruments \\
\hline
\endhead
56123.6 & -8.8 & 15.024(0.022) & $\cdots$ & 14.052(0.025) & $\cdots$ & ST10 \\
56124.7 & -7.7 & 14.791(0.030) & 14.740(0.023) & 14.696(0.025) & 14.655(0.036) & ST10 \\
56125.6 & -6.8 & 14.708(0.020) & 14.659(0.019) & 14.631(0.028) & 14.594(0.029) & ST10 \\
56126.6 & -5.8 & 14.641(0.025) & 14.413(0.011) & 14.541(0.036) & 14.532(0.028) & ST10 \\
56131.6 & -0.8 & 14.442(0.022) & 14.392(0.021) & 14.356(0.065) & 14.279(0.137) & ST10 \\
56132.7 & 0.3 & 14.578(0.058) & 14.364(0.023) & 14.363(0.050) & 14.364(0.036) & ST10 \\
56133.6 & 1.2 & $\cdots$ & 14.381(0.011) & 14.286(0.095) & 14.339(0.027) & ST10 \\
56134.7 & 2.3 & $\cdots$ & 14.369(0.021) & 14.306(0.030) & 14.344(0.040) & ST10 \\
56136.6 & 4.2 & $\cdots$ & 14.392(0.038) & 14.346(0.039) & $\cdots$ & Andor \\
56137.7 & 5.3 & $\cdots$ & 14.373(0.075) & $\cdots$ & $\cdots$ & Andor \\
56141.7 & 9.3 & $\cdots$ & 14.431(0.033) & 14.408(0.042) & $\cdots$ & Andor \\
56142.6 & 10.2 & 14.782(0.030) & 14.495(0.017) & 14.402(0.026) & 14.354(0.028) & Andor \\
56143.6 & 11.2 & 14.892(0.042) & 14.510(0.025) & 14.597(0.171) & $\cdots$ & Andor \\
56146.6 & 14.2 & 15.236(0.023) & 14.627(0.016) & 14.510(0.030) & 14.368(0.026) & Andor \\
56147.7 & 15.3 & 15.288(0.027) & 14.679(0.016) & 14.553(0.030) & 14.378(0.034) & Andor \\
56154.6 & 22.2 & 16.198(0.035) & 15.069(0.031) & 14.756(0.028) & 14.367(0.028) & Andor \\
56155.6 & 23.2 & $\cdots$ & 15.126(0.024) & 14.782(0.040) & 14.378(0.028) & ST10 \\
56156.6 & 24.2 & $\cdots$ & $\cdots$ & 14.780(0.026) & 14.414(0.026) & Andor \\
56158.6 & 26.2 & $\cdots$ & 15.293(0.025) & 14.895(0.029) & 14.442(0.037) & ST10 \\
56159.6 & 27.2 & $\cdots$ & 15.240(0.046) & $\cdots$ & $\cdots$ & ST10 \\
56161.6 & 29.2 & $\cdots$ & $\cdots$ & $\cdots$ & 14.415(0.278) & ST10 \\
56164.6 & 32.2 & $\cdots$ & $\cdots$ & 15.058(0.027) & $\cdots$ & Andor \\
56165.6 & 33.2 & $\cdots$ & $\cdots$ & 15.093(0.026) & 14.644(0.027) & Andor \\
56174.6 & 42.2 & $\cdots$ & $\cdots$ & 15.404(0.050) & $\cdots$ & Andor \\
\hline
\end{longtable}{}
\end{center}

\begin{center}
\scriptsize
\begin{longtable}{lllllll}
    \caption{HOWPol photometry of SN 2012dn in the $UBVRI$ bands} \\
\hline
\hline
MJD     & Phase & $B$         & $V$     & $R$ & $I$ & $z'+Y$ \\
\hline
\endhead
56131.6 & -0.8 & 14.425(0.036) & 14.303(0.029) & 14.314(0.041) & 14.341(0.048) & 14.777(0.068) \\
56132.6 & 0.2 & 14.439(0.028) & 14.316(0.024) & 14.300(0.041) & 14.323(0.038) & 14.719(0.062) \\
56133.7 & 1.3 & 14.461(0.028) & 14.305(0.023) & 14.274(0.035) & 14.302(0.038) & 14.677(0.065) \\
56134.7 & 2.3 & 14.481(0.050) & 14.337(0.029) & 14.278(0.056) & 14.322(0.053) & 14.668(0.069) \\
56135.7 & 3.3 & $\cdots$ & 14.526(0.126) & 14.298(0.073) & 14.399(0.085) & 14.746(0.078) \\
56137.5 & 5.1 & 14.623(0.089) & 14.389(0.091) & 14.291(0.039) & 14.294(0.041) & 14.631(0.064) \\
56138.6 & 6.2 & 14.668(0.056) & 14.366(0.034) & 14.349(0.033) & 14.294(0.045) & 14.680(0.086) \\
56139.6 & 7.2 & 14.687(0.045) & 14.382(0.040) & 14.286(0.034) & 14.334(0.055) & 14.581(0.078) \\
56140.6 & 8.2 & 14.792(0.047) & 14.396(0.029) & 14.371(0.038) & 14.400(0.045) & 14.652(0.070) \\
56141.7 & 9.3 & 14.816(0.037) & 14.481(0.027) & 14.370(0.044) & 14.383(0.055) & 14.643(0.062) \\
56142.6 & 10.2 & 14.926(0.060) & 14.424(0.050) & 14.369(0.072) & 14.347(0.117) & 14.616(0.081) \\
56143.6 & 11.2 & 15.002(0.033) & 14.442(0.033) & 14.417(0.034) & 14.355(0.042) & 14.605(0.065) \\
56146.5 & 14.1 & $\cdots$ & 14.567(0.022) & 14.487(0.026) & 14.368(0.025) & 14.515(0.130) \\
56149.5 & 17.1 & 15.644(0.040) & 14.696(0.047) & 14.553(0.042) & 14.383(0.047) & 14.540(0.070) \\
56150.5 & 18.1 & 15.868(0.074) & 14.827(0.050) & 14.713(0.049) & 14.577(0.058) & 14.680(0.129) \\
56155.5 & 23.1 & 16.355(0.043) & 15.009(0.030) & 14.776(0.049) & 14.409(0.055) & 14.530(0.074) \\
56156.5 & 24.1 & 16.393(0.058) & 15.100(0.074) & 14.836(0.068) & 14.536(0.056) & 14.574(0.078) \\
56158.5 & 26.1 & 16.704(0.049) & 15.271(0.033) & 14.984(0.074) & 14.555(0.042) & 14.656(0.076) \\
56159.6 & 27.2 & 16.745(0.040) & 15.267(0.043) & 14.915(0.033) & 14.508(0.042) & 14.607(0.070) \\
56160.5 & 28.1 & 16.858(0.048) & 15.302(0.030) & 14.994(0.038) & 14.565(0.038) & 14.662(0.069) \\
56161.5 & 29.1 & 16.845(0.066) & 15.345(0.047) & 15.028(0.049) & 14.624(0.036) & 14.594(0.107) \\
56162.5 & 30.1 & $\cdots$ & 15.525(0.033) & $\cdots$ & $\cdots$ & $\cdots$ \\
56163.6 & 31.2 & 16.986(0.066) & 15.468(0.050) & 15.142(0.053) & 14.602(0.057) & 14.687(0.090) \\
56164.6 & 32.2 & 17.057(0.061) & 15.487(0.053) & 15.145(0.029) & 14.644(0.041) & 14.737(0.101) \\
56165.6 & 33.2 & 17.120(0.059) & 15.495(0.019) & 15.152(0.031) & 14.689(0.027) & 14.760(0.080) \\
56170.5 & 38.1 & $\cdots$ & 15.717(0.073) & 15.393(0.061) & 14.888(0.045) & 14.893(0.071) \\
56172.5 & 40.1 & 17.430(0.043) & 15.841(0.033) & 15.498(0.043) & 14.926(0.046) & 15.001(0.066) \\
56182.5 & 50.1 & 17.649(0.058) & 16.119(0.037) & 15.900(0.044) & 15.335(0.051) & 15.336(0.071) \\
56184.5 & 52.1 & 17.753(0.086) & 16.207(0.044) & 16.001(0.043) & 15.407(0.035) & 15.426(0.067) \\
56187.5 & 55.1 & 17.702(0.066) & 16.279(0.021) & 16.062(0.046) & 15.682(0.104) & $\cdots$ \\
56188.6 & 56.2 & 17.873(0.089) & 16.276(0.031) & 16.110(0.044) & 15.540(0.044) & $\cdots$ \\
56193.5 & 61.1 & 17.937(0.098) & 16.490(0.042) & 16.348(0.039) & 15.781(0.040) & 15.766(0.065) \\
56196.5 & 64.1 & $\cdots$ & 16.737(0.127) & $\cdots$ & 15.921(0.086) & 15.845(0.097) \\
56202.5 & 70.1 & 17.902(0.066) & 17.118(0.051) & $\cdots$ & $\cdots$ & 16.175(0.063) \\
56203.5 & 71.1 & 18.019(0.092) & 17.166(0.037) & $\cdots$ & $\cdots$ & $\cdots$ \\
56207.5 & 75.1 & 18.058(0.074) & 16.949(0.038) & 16.800(0.047) & 16.308(0.047) & 16.469(0.071) \\
56209.5 & 77.1 & $\cdots$ & 17.191(0.044) & 17.059(0.178) & 16.338(0.050) & 16.491(0.076) \\
56211.5 & 79.1 & $\cdots$ & $\cdots$ & 16.777(0.042) & $\cdots$ & $\cdots$ \\
56212.4 & 80.0 & $\cdots$ & 17.005(0.028) & 16.895(0.029) & 16.417(0.039) & 16.644(0.069) \\
56213.4 & 81.0 & $\cdots$ & 17.207(0.057) & 17.186(0.112) & 16.585(0.032) & $\cdots$ \\
56214.5 & 82.1 & $\cdots$ & $\cdots$ & 16.756(0.050) & 16.607(0.054) & 16.853(0.073) \\
56215.5 & 83.1 & $\cdots$ & 17.229(0.051) & 17.050(0.039) & 16.582(0.049) & 16.671(0.073) \\
56216.5 & 84.1 & $\cdots$ & $\cdots$ & 16.880(0.037) & 16.550(0.053) & 16.696(0.078) \\
56220.5 & 88.1 & $\cdots$ & 17.454(0.058) & 17.239(0.036) & 16.805(0.047) & 17.059(0.087) \\
56221.4 & 89.0 & $\cdots$ & 17.328(0.064) & 17.096(0.061) & 16.636(0.050) & 16.773(0.082) \\
56223.4 & 91.0 & $\cdots$ & 17.704(0.233) & 17.369(0.113) & 17.069(0.093) & 17.091(0.112) \\
56224.4 & 92.0 & $\cdots$ & 17.175(0.071) & 17.085(0.083) & 16.500(0.093) & 16.894(0.118) \\
56225.4 & 93.0 & $\cdots$ & 17.259(0.084) & 17.164(0.076) & 16.890(0.076) & $\cdots$ \\
56226.4 & 94.0 & $\cdots$ & 17.354(0.076) & 17.183(0.090) & $\cdots$ & $\cdots$ \\
56228.4 & 96.0 & $\cdots$ & $\cdots$ & 16.904(0.086) & 16.676(0.068) & 16.834(0.089) \\
56228.4 & 96.0 & $\cdots$ & $\cdots$ & $\cdots$ & 16.676(0.076) & 16.605(0.532) \\
56233.4 & 101.0 & $\cdots$ & $\cdots$ & $\cdots$ & 17.107(0.062) & $\cdots$ \\
\hline
\end{longtable}{}
\end{center}

\begin{center}
\scriptsize
\begin{longtable}{llllll}
    \caption{NIR-band photometry of SN 2012dn} \\
\hline
\hline
MJD     & Phase &$J$         & $H$         & $K_{s}$     & Instruments \\
\hline
\endhead
56120.8 & -11.6 & 15.09(0.04) & 15.09(0.04) & 15.04(0.09) & IRSF \\
56125.1 & -7.3 & 14.62(0.03) & 14.70(0.03) & 14.57(0.06) & IRSF \\
56125.7 & -6.7 & 14.67(0.19) & 14.72(0.02) & $\cdots$ & NIC \\
56126.1 & -6.3 & 14.56(0.03) & 14.58(0.03) & 14.56(0.07) & IRSF \\
56127.1 & -5.3 & 14.50(0.03) & 14.59(0.03) & 14.58(0.08) & IRSF \\
56129.6 & -2.8 & 14.68(0.02) & 14.82(0.04) & $\cdots$ & IR Cam\\
56132.0 & -0.4 & 14.34(0.03) & 14.44(0.03) & 14.24(0.05) & IRSF \\
56132.7 & 0.3 & 14.34(0.03) & $\cdots$ & $\cdots$ & IR Cam\\
56133.0 & 0.6 & 14.43(0.03) & 14.46(0.03) & 14.19(0.05) & IRSF \\
56133.6 & 1.2 & 14.55(0.05) & 14.63(0.05) & $\cdots$ & IR Cam\\
56133.7 & 1.3 & 14.39(0.12) & 14.64(0.09) & $\cdots$ & NIC \\
56133.7 & 1.3 & $\cdots$ & $\cdots$ & 14.36(0.20) & NIC \\
56134.0 & 1.6 & 14.42(0.03) & 14.46(0.03) & 14.37(0.08) & IRSF \\
56134.6 & 2.2 & 14.46(0.03) & 14.51(0.02) & 14.37(0.06) & NIC \\
56135.6 & 3.2 & 14.36(0.17) & 14.50(0.07) & 14.45(0.10) & NIC \\
56136.0 & 3.6 & 14.38(0.03) & 14.38(0.03) & 14.21(0.05) & IRSF \\
56136.6 & 4.2 & 14.55(0.07) & 14.52(0.02) & 14.43(0.08) & NIC \\
56137.0 & 4.6 & 14.43(0.03) & 14.43(0.03) & 14.37(0.05) & IRSF \\
56137.6 & 5.2 & 14.51(0.04) & 14.48(0.03) & 14.36(0.05) & NIC \\
56138.0 & 5.6 & 14.39(0.03) & 14.34(0.03) & 14.16(0.05) & IRSF \\
56138.6 & 6.2 & 14.56(0.01) & 14.49(0.02) & 14.34(0.04) & NIC \\
56139.0 & 6.6 & 14.40(0.03) & 14.34(0.03) & 14.22(0.05) & IRSF \\
56139.6 & 7.2 & 14.52(0.03) & 14.46(0.01) & 14.33(0.03) & NIC \\
56140.6 & 8.2 & 14.61(0.12) & 14.47(0.04) & 14.36(0.05) & NIC \\
56141.1 & 8.7 & 14.48(0.03) & 14.27(0.03) & 14.13(0.05) & IRSF \\
56141.6 & 9.2 & 14.61(0.02) & 14.44(0.02) & 14.26(0.04) & NIC \\
56142.1 & 9.7 & 14.51(0.03) & 14.38(0.03) & 14.25(0.06) & IRSF \\
56143.1 & 10.7 & 14.60(0.03) & 14.41(0.03) & 14.10(0.05) & IRSF \\
56143.6 & 11.2 & 14.71(0.02) & 14.42(0.01) & 14.27(0.05) & NIC \\
56145.6 & 13.2 & 14.80(0.03) & 14.42(0.02) & 14.29(0.02) & NIC \\
56146.7 & 14.3 & 14.80(0.06) & $\cdots$ & $\cdots$ & IR Cam\\
56149.1 & 16.7 & 14.73(0.04) & 14.32(0.03) & 14.16(0.06) & IRSF \\
56153.0 & 20.6 & 14.72(0.04) & 14.26(0.03) & 14.25(0.06) & IRSF \\
56156.6 & 24.2 & 14.85(0.04) & $\cdots$ & $\cdots$ & IR Cam\\
56159.6 & 27.2 & 14.74(0.04) & 14.31(0.01) & 14.22(0.04) & NIC \\
56161.0 & 28.6 & 14.60(0.03) & 14.22(0.03) & 14.13(0.04) & IRSF \\
56163.0 & 30.6 & 14.55(0.03) & 14.21(0.03) & 14.12(0.05) & IRSF \\
56164.6 & 32.2 & 14.73(0.00) & 14.29(0.01) & 14.21(0.04) & NIC \\
56165.6 & 33.2 & 14.73(0.01) & 14.26(0.01) & 14.13(0.05) & NIC \\
56166.9 & 34.5 & 14.64(0.03) & 14.19(0.03) & 14.05(0.05) & IRSF \\
56174.6 & 42.2 & 14.98(0.04) & 14.32(0.01) & 14.09(0.05) & NIC \\
56174.8 & 42.4 & 14.82(0.03) & 14.27(0.02) & 13.94(0.04) & IRSF \\
56175.8 & 43.4 & 14.87(0.04) & 14.19(0.03) & 13.89(0.04) & IRSF \\
56176.8 & 44.4 & 14.94(0.04) & 14.24(0.03) & 14.00(0.05) & IRSF \\
56177.7 & 45.3 & 15.01(0.04) & 14.23(0.03) & 13.94(0.05) & IRSF \\
56178.7 & 46.3 & 14.99(0.04) & 14.23(0.03) & 13.97(0.05) & IRSF \\
56180.7 & 48.3 & 15.05(0.04) & 14.27(0.03) & 13.85(0.04) & IRSF \\
56181.7 & 49.3 & 15.07(0.04) & 14.28(0.03) & 13.86(0.05) & IRSF \\
56182.6 & 50.2 & 15.55(0.05) & $\cdots$ & $\cdots$ & IR Cam\\
56182.7 & 50.3 & 15.25(0.05) & 14.27(0.03) & 13.80(0.05) & IRSF \\
56183.5 & 51.1 & 15.68(0.05) & 14.39(0.04) & 13.98(0.06) & IR Cam\\
56184.7 & 52.3 & 15.24(0.05) & 14.33(0.03) & 13.86(0.04) & IRSF \\
56185.7 & 53.3 & 15.22(0.05) & 14.27(0.03) & 13.83(0.05) & IRSF \\
56186.7 & 54.3 & 15.10(0.05) & 14.29(0.03) & 13.83(0.06) & IRSF \\
56189.8 & 57.4 & 15.38(0.04) & 14.38(0.03) & 13.78(0.03) & IRSF \\
56190.7 & 58.3 & 15.43(0.05) & 14.39(0.02) & 13.78(0.03) & IRSF \\
56192.7 & 60.3 & 15.43(0.05) & 14.41(0.03) & 13.81(0.03) & IRSF \\
56193.5 & 61.1 & 15.54(0.05) & $\cdots$ & $\cdots$ & IR Cam\\
56194.7 & 62.3 & 15.35(0.07) & 14.40(0.03) & 13.88(0.04) & IRSF \\
56195.7 & 63.3 & 15.59(0.06) & 14.39(0.03) & 13.80(0.05) & IRSF \\
56196.6 & 64.2 & 15.85(0.06) & 14.60(0.03) & 13.84(0.05) & NIC \\
56196.7 & 64.3 & 15.69(0.06) & 14.52(0.03) & 13.82(0.05) & IRSF \\
56199.7 & 67.3 & 15.62(0.06) & 14.51(0.03) & 13.83(0.04) & IRSF \\
56200.7 & 68.3 & 15.87(0.09) & 14.50(0.04) & 13.77(0.05) & IRSF \\
56201.5 & 69.1 & $\cdots$ & 14.44(0.07) & 13.83(0.07) & IR Cam\\
56201.7 & 69.3 & 15.60(0.06) & 14.53(0.04) & 13.71(0.04) & IRSF \\
56202.5 & 70.1 & $\cdots$ & 14.80(0.05) & 13.83(0.05) & IR Cam\\
56202.5 & 70.1 & 15.92(0.11) & 14.70(0.03) & 14.00(0.04) & NIC \\
56202.7 & 70.3 & 15.97(0.07) & 14.69(0.04) & 13.87(0.05) & IRSF \\
56203.8 & 71.4 & 15.77(0.05) & 14.60(0.03) & 13.84(0.03) & IRSF \\
56204.5 & 72.1 & 16.04(0.18) & 14.91(0.04) & 14.09(0.09) & NIC \\
56204.5 & 72.1 & $\cdots$ & 14.66(0.08) & 13.78(0.06) & IR Cam\\
56205.5 & 73.1 & $\cdots$ & $\cdots$ & 13.96(0.06) & IR Cam\\
56205.7 & 73.3 & 15.79(0.07) & 14.62(0.04) & 13.80(0.04) & IRSF \\
56207.5 & 75.1 & 16.12(0.27) & 14.83(0.03) & 14.01(0.04) & NIC \\
56208.5 & 76.1 & $\cdots$ & $\cdots$ & 14.07(0.04) & IR Cam\\
56208.8 & 76.4 & 15.94(0.07) & 14.68(0.04) & 13.94(0.05) & IRSF \\
56210.4 & 78.0 & $\cdots$ & 14.95(0.09) & $\cdots$ & IR Cam\\
56211.5 & 79.1 & $\cdots$ & 14.81(0.05) & $\cdots$ & IR Cam\\
56211.8 & 79.4 & 16.22(0.08) & 14.86(0.04) & 13.91(0.04) & IRSF \\
56212.5 & 80.1 & 16.43(0.22) & 15.00(0.10) & 14.13(0.07) & NIC \\
56212.5 & 80.1 & $\cdots$ & $\cdots$ & 14.08(0.10) & IR Cam\\
56212.8 & 80.4 & 16.04(0.08) & 14.84(0.05) & 13.95(0.05) & IRSF \\
56213.9 & 81.5 & 16.13(0.07) & 14.87(0.04) & 14.01(0.05) & IRSF \\
56214.5 & 82.1 & 16.53(0.21) & 15.19(0.12) & 14.24(0.09) & NIC \\
56216.4 & 84.0 & $\cdots$ & 15.26(0.04) & 14.34(0.07) & NIC \\
56220.4 & 88.0 & $\cdots$ & 15.13(0.11) & 14.36(0.12) & IR Cam\\
56223.4 & 91.0 & $\cdots$ & 15.08(0.08) & 14.33(0.08) & IR Cam\\
56225.8 & 93.4 & 16.71(0.11) & 15.17(0.05) & 14.26(0.06) & IRSF \\
56226.4 & 94.0 & 16.68(0.25) & 15.60(0.05) & 14.48(0.07) & NIC \\
56226.8 & 94.4 & 16.72(0.13) & 15.29(0.06) & 14.16(0.06) & IRSF \\
56227.8 & 95.4 & 16.59(0.17) & 15.18(0.08) & 14.21(0.10) & IRSF \\
56228.4 & 96.0 & $\cdots$ & $\cdots$ & 14.24(0.06) & IR Cam\\
56228.7 & 96.3 & 16.71(0.16) & 15.40(0.08) & 14.11(0.07) & IRSF \\
56231.8 & 99.4 & 16.81(0.10) & 15.40(0.05) & 14.26(0.04) & IRSF \\
56232.8 & 100.4 & 16.92(0.11) & 15.50(0.05) & 14.33(0.05) & IRSF \\
56233.8 & 101.4 & 16.95(0.14) & 15.59(0.07) & 14.35(0.06) & IRSF \\
56237.4 & 105.0 & $\cdots$ & 16.31(0.25) & 15.13(0.37) & NIC \\
56238.7 & 106.3 & 17.50(0.25) & 15.72(0.07) & 14.40(0.06) & IRSF \\
56240.7 & 108.3 & 16.88(0.09) & 15.62(0.05) & 14.50(0.05) & IRSF \\
56241.7 & 109.3 & 17.03(0.11) & 15.73(0.06) & 14.53(0.05) & IRSF \\
56242.7 & 110.3 & 17.08(0.16) & 15.83(0.09) & 14.45(0.06) & IRSF \\
56242.7 & 110.3 & 17.08(0.16) & 15.83(0.09) & 14.45(0.06) & IRSF \\
56243.8 & 111.4 & 17.18(0.11) & 15.73(0.06) & 14.48(0.04) & IRSF \\
56243.8 & 111.4 & 17.18(0.11) & 15.73(0.06) & 14.48(0.04) & IRSF \\
56244.8 & 112.4 & 17.24(0.13) & 15.74(0.06) & 14.57(0.05) & IRSF \\
56244.8 & 112.4 & 17.24(0.13) & 15.74(0.06) & 14.57(0.05) & IRSF \\
56245.7 & 113.3 & 17.63(0.20) & 15.80(0.05) & 14.68(0.05) & IRSF \\
56246.7 & 114.3 & 17.11(0.14) & 15.88(0.06) & 14.55(0.05) & IRSF \\
56247.7 & 115.3 & 17.01(0.11) & 15.81(0.05) & 14.76(0.05) & IRSF \\
56248.7 & 116.3 & 17.24(0.14) & 16.12(0.07) & 14.78(0.06) & IRSF \\
56249.8 & 117.4 & 17.24(0.12) & 15.89(0.06) & 14.70(0.05) & IRSF \\
56250.7 & 118.3 & 17.24(0.12) & 16.01(0.06) & 14.75(0.05) & IRSF \\
56251.7 & 119.3 & 17.83(0.19) & 16.14(0.07) & 14.94(0.05) & IRSF \\
56257.8 & 125.4 & $\cdots$ & 15.96(0.09) & 15.13(0.13) & IRSF \\
56259.8 & 127.4 & $\cdots$ & 16.70(0.25) & 15.16(0.18) & IRSF \\
56264.8 & 132.4 & $\cdots$ & 16.38(0.12) & 15.11(0.10) & IRSF \\
56267.8 & 135.4 & $\cdots$ & 16.77(0.13) & 15.24(0.08) & IRSF \\
56268.8 & 136.4 & $\cdots$ & 16.51(0.11) & 15.23(0.08) & IRSF \\
56269.8 & 137.4 & $\cdots$ & 16.34(0.12) & 15.42(0.12) & IRSF \\
56272.8 & 140.4 & $\cdots$ & 16.80(0.17) & 15.32(0.10) & IRSF \\
56348.1 & 215.7 & $>19.05$ & $>20.82$ & $>19.39$ & IRSF \\

\hline
\end{longtable}{}
\end{center}

\begin{center}
\scriptsize
\begin{longtable}{llllll}
    \caption{Log of spectroscopic observations of SN 2012dn} \\
\hline
\hline
MJD     & Phase & Coverage    & Resolutions & Instruments \\
\hline
\endhead
56131.6 & -0.8 & 4500-9000\AA & 15.0\AA & HOWPol \\
56132.6 & 0.2 & 4500-9000\AA & 15.0\AA & HOWPol \\
56133.7 & 1.3 & 4500-9000\AA & 15.0\AA & HOWPol \\
56134.7 & 2.3 & 4500-9000\AA & 15.0\AA & HOWPol \\
56137.5 & 5.1 & 4500-9000\AA & 15.0\AA & HOWPol \\
56138.6 & 6.2 & 4500-9000\AA & 15.0\AA & HOWPol \\
56139.6 & 7.2 & 4500-9000\AA & 15.0\AA & HOWPol \\
56140.6 & 8.2 & 4500-9000\AA & 15.0\AA & HOWPol \\
56141.6 & 9.2 & 4500-9000\AA & 15.0\AA & HOWPol \\
56142.6 & 10.2 & 4500-9000\AA & 15.0\AA & HOWPol \\
56143.6 & 11.2 & 4500-9000\AA & 15.0\AA & HOWPol \\
56149.6 & 17.2 & 4500-9000\AA & 15.0\AA & HOWPol \\
56150.5 & 18.1 & 4500-9000\AA & 15.0\AA & HOWPol \\
56155.5 & 23.1 & 4500-9000\AA & 15.0\AA & HOWPol \\
56156.5 & 24.1 & 4500-9000\AA & 15.0\AA & HOWPol \\
56158.5 & 26.1 & 4500-9000\AA & 15.0\AA & HOWPol \\
56159.6 & 27.2 & 4500-9000\AA & 15.0\AA & HOWPol \\
56160.5 & 28.1 & 4500-9000\AA & 15.0\AA & HOWPol \\
56163.6 & 31.2 & 4500-9000\AA & 15.0\AA & HOWPol \\
56164.6 & 32.2 & 4500-9000\AA & 15.0\AA & HOWPol \\
56165.6 & 33.2 & 4500-9000\AA & 15.0\AA & HOWPol \\
56170.5 & 38.1 & 4500-9000\AA & 15.0\AA & HOWPol \\
56172.5 & 40.1 & 4500-9000\AA & 15.0\AA & HOWPol \\
56182.5 & 50.1 & 4500-9000\AA & 15.0\AA & HOWPol \\
56183.5 & 51.1 & 4500-9000\AA & 15.0\AA & HOWPol \\
56184.5 & 52.1 & 4500-9000\AA & 15.0\AA & HOWPol \\
56193.5 & 61.1 & 4500-9000\AA & 15.0\AA & HOWPol \\
56196.5 & 64.1 & 4500-9000\AA & 15.0\AA & HOWPol \\
56197.4 & 65.0 & 4500-9000\AA & 15.0\AA & HOWPol \\
\hline
\end{longtable}{}
\end{center}


\begin{thebibliography}{115}
\expandafter\ifx\csname natexlab\endcsname\relax\def\natexlab#1{#1}\fi

\bibitem[{{Altavilla} {et~al.}(2004){Altavilla}, {Fiorentino}, {Marconi},
  {Musella}, {Cappellaro}, {Barbon}, {Benetti}, {Pastorello}, {Riello},
  {Turatto}, \& {Zampieri}}]{Altavilla2004}
{Altavilla}, G., {et~al.} 2004, \mnras, 349, 1344

\bibitem[{{Andrews} {et~al.}(2011){Andrews}, {Clayton}, {Wesson}, {Sugerman},
  {Barlow}, {Clem}, {Ercolano}, {Fabbri}, {Gallagher}, {Landolt}, {Meixner},
  {Otsuka}, {Riebel}, \& {Welch}}]{Andrews2011}
{Andrews}, J.~E., {et~al.} 2011, \aj, 142, 45

\bibitem[{{Arnett}(1982)}]{Arnett1982}
{Arnett}, W.~D. 1982, \apj, 253, 785

\bibitem[{{Bessell}(1990)}]{Bessell1990}
{Bessell}, M.~S. 1990, \pasp, 102, 1181

\bibitem[{{Bock} {et~al.}(2012){Bock}, {Parrent}, \& {Howell}}]{Bock2012}
{Bock}, G., {Parrent}, J.~T., \& {Howell}, D.~A. 2012, Central Bureau
  Electronic Telegrams, 3174

\bibitem[{{Bode} \& {Evans}(1980)}]{Bode1980}
{Bode}, M.~F., \& {Evans}, A. 1980, \mnras, 193, 21P

\bibitem[{{Bode} {et~al.}(2007){Bode}, {Harman}, {O'Brien}, {Bond},
  {Starrfield}, {Darnley}, {Evans}, \& {Eyres}}]{Bode2007}
{Bode}, M.~F., {Harman}, D.~J., {O'Brien}, T.~J., {Bond}, H.~E., {Starrfield},
  S., {Darnley}, M.~J., {Evans}, A., \& {Eyres}, S.~P.~S. 2007, \apjl, 665, L63

\bibitem[{{Branch} {et~al.}(1993){Branch}, {Fisher}, \& {Nugent}}]{Branch1993}
{Branch}, D., {Fisher}, A., \& {Nugent}, P. 1993, \aj, 106, 2383

\bibitem[{{Brown}(2014)}]{Brown2014b}
{Brown}, P.~J. 2014, \apjl, 796, L18

\bibitem[{{Brown} {et~al.}(2014){Brown}, {Kuin}, {Scalzo}, {Smitka}, {de
  Pasquale}, {Holland}, {Krisciunas}, {Milne}, \& {Wang}}]{Brown2014a}
{Brown}, P.~J., {et~al.} 2014, \apj, 787, 29

\bibitem[{{Cappellaro} {et~al.}(2001){Cappellaro}, {Patat}, {Mazzali},
  {Benetti}, {Danziger}, {Pastorello}, {Rizzi}, {Salvo}, \&
  {Turatto}}]{Cappellaro2001}
{Cappellaro}, E., {et~al.} 2001, \apjl, 549, L215

\bibitem[{{Chakradhari} {et~al.}(2014){Chakradhari}, {Sahu}, {Srivastav}, \&
  {Anupama}}]{Chakradhari2014}
{Chakradhari}, N.~K., {Sahu}, D.~K., {Srivastav}, S., \& {Anupama}, G.~C. 2014,
  \mnras, 443, 1663

\bibitem[{{Chevalier}(1986)}]{Chevalier1986}
{Chevalier}, R.~A. 1986, \apj, 308, 225

\bibitem[{{Crotts}(2015)}]{Crotts2015}
{Crotts}, A.~P.~S. 2015, \apjl, 804, L37

\bibitem[{{Dhawan} {et~al.}(2015){Dhawan}, {Leibundgut}, {Spyromilio}, \&
  {Maguire}}]{Dhawan2015}
{Dhawan}, S., {Leibundgut}, B., {Spyromilio}, J., \& {Maguire}, K. 2015,
  \mnras, 448, 1345

\bibitem[{{Di Carlo} {et~al.}(2008){Di Carlo}, {Corsi}, {Arkharov}, {Massi},
  {Larionov}, {Efimova}, {Dolci}, {Napoleone}, \& {Di Paola}}]{DiCarlo2008}
{Di Carlo}, E., {et~al.} 2008, \apj, 684, 471

\bibitem[{{Dilday} {et~al.}(2012){Dilday}, {Howell}, {Cenko}, {Silverman},
  {Nugent}, {Sullivan}, {Ben-Ami}, {Bildsten}, {Bolte}, {Endl}, {Filippenko},
  {Gnat}, {Horesh}, {Hsiao}, {Kasliwal}, {Kirkman}, {Maguire}, {Marcy},
  {Moore}, {Pan}, {Parrent}, {Podsiadlowski}, {Quimby}, {Sternberg}, {Suzuki},
  {Tytler}, {Xu}, {Bloom}, {Gal-Yam}, {Hook}, {Kulkarni}, {Law}, {Ofek},
  {Polishook}, \& {Poznanski}}]{Dilday2012}
{Dilday}, B., {et~al.} 2012, Science, 337, 942

\bibitem[{{Drozdov} {et~al.}(2015){Drozdov}, {Leising}, {Milne}, {Pearcy},
  {Riess}, {Macri}, {Bryngelson}, \& {Garnavich}}]{Drozdov2015}
{Drozdov}, D., {Leising}, M.~D., {Milne}, P.~A., {Pearcy}, J., {Riess}, A.~G.,
  {Macri}, L.~M., {Bryngelson}, G.~L., \& {Garnavich}, P.~M. 2015, \apj, 805,
  71

\bibitem[{{Dwek}(1983)}]{Dwek1983b}
{Dwek}, E. 1983, \apj, 274, 175

\bibitem[{{Dwek} {et~al.}(1983){Dwek}, {A'Hearn}, {Becklin}, {Brown}, {Capps},
  {Dinerstein}, {Gatley}, {Morrison}, {Telesco}, {Tokunaga}, {Werner}, \&
  {Wynn-Williams}}]{Dwek1983a}
{Dwek}, E., {et~al.} 1983, \apj, 274, 168

\bibitem[{{Folatelli} {et~al.}(2010){Folatelli}, {Phillips}, {Burns},
  {Contreras}, {Hamuy}, {Freedman}, {Persson}, {Stritzinger}, {Suntzeff},
  {Krisciunas}, {Boldt}, {Gonz{\'a}lez}, {Krzeminski}, {Morrell}, {Roth},
  {Salgado}, {Madore}, {Murphy}, {Wyatt}, {Li}, {Filippenko}, \&
  {Miller}}]{Folatelli2010}
{Folatelli}, G., {et~al.} 2010, \aj, 139, 120

\bibitem[{{Fox} {et~al.}(2009){Fox}, {Skrutskie}, {Chevalier}, {Kanneganti},
  {Park}, {Wilson}, {Nelson}, {Amirhadji}, {Crump}, {Hoeft}, {Provence},
  {Sargeant}, {Sop}, {Tea}, {Thomas}, \& {Woolard}}]{Fox2009}
{Fox}, O., {et~al.} 2009, \apj, 691, 650

\bibitem[{{Fox} {et~al.}(2013){Fox}, {Filippenko}, {Skrutskie}, {Silverman},
  {Ganeshalingam}, {Cenko}, \& {Clubb}}]{Fox2013}
{Fox}, O.~D., {Filippenko}, A.~V., {Skrutskie}, M.~F., {Silverman}, J.~M.,
  {Ganeshalingam}, M., {Cenko}, S.~B., \& {Clubb}, K.~I. 2013, \aj, 146, 2

\bibitem[{{Fox} {et~al.}(2016){Fox}, {Johansson}, {Kasliwal}, {Andrews},
  {Bally}, {Bond}, {Boyer}, {Gehrz}, {Helou}, {Hsiao}, {Masci},
  {Parthasarathy}, {Smith}, {Tinyanont}, \& {Van Dyk}}]{Fox2016}
{Fox}, O.~D., {et~al.} 2016, \apjl, 816, L13

\bibitem[{{Friedman} {et~al.}(2015){Friedman}, {Wood-Vasey}, {Marion},
  {Challis}, {Mandel}, {Bloom}, {Modjaz}, {Narayan}, {Hicken}, {Foley},
  {Klein}, {Starr}, {Morgan}, {Rest}, {Blake}, {Miller}, {Falco}, {Wyatt},
  {Mink}, {Skrutskie}, \& {Kirshner}}]{Friedman2015}
{Friedman}, A.~S., {et~al.} 2015, \apjs, 220, 9

\bibitem[{{Fukugita} {et~al.}(1996){Fukugita}, {Ichikawa}, {Gunn}, {Doi},
  {Shimasaku}, \& {Schneider}}]{Fukugita1996}
{Fukugita}, M., {Ichikawa}, T., {Gunn}, J.~E., {Doi}, M., {Shimasaku}, K., \&
  {Schneider}, D.~P. 1996, \aj, 111, 1748

\bibitem[{{Gall} {et~al.}(2014){Gall}, {Hjorth}, {Watson}, {Dwek}, {Maund},
  {Fox}, {Leloudas}, {Malesani}, \& {Day-Jones}}]{Gall2014}
{Gall}, C., {et~al.} 2014, \nat, 511, 326

\bibitem[{{Gerardy} {et~al.}(2000){Gerardy}, {Fesen}, {H{\"o}flich}, \&
  {Wheeler}}]{Gerardy2000}
{Gerardy}, C.~L., {Fesen}, R.~A., {H{\"o}flich}, P., \& {Wheeler}, J.~C. 2000,
  \aj, 119, 2968

\bibitem[{{Hachinger} {et~al.}(2012){Hachinger}, {Mazzali}, {Taubenberger},
  {Fink}, {Pakmor}, {Hillebrandt}, \& {Seitenzahl}}]{Hachinger2012}
{Hachinger}, S., {Mazzali}, P.~A., {Taubenberger}, S., {Fink}, M., {Pakmor},
  R., {Hillebrandt}, W., \& {Seitenzahl}, I.~R. 2012, \mnras, 427, 2057

\bibitem[{{Hachisu} \& {Kato}(2001)}]{Hachisu2001}
{Hachisu}, I., \& {Kato}, M. 2001, \apj, 558, 323

\bibitem[{{Hachisu} {et~al.}(2000){Hachisu}, {Kato}, {Kato}, \&
  {Matsumoto}}]{Hachisu2000}
{Hachisu}, I., {Kato}, M., {Kato}, T., \& {Matsumoto}, K. 2000, \apjl, 528, L97

\bibitem[{{Hachisu} {et~al.}(1999){Hachisu}, {Kato}, {Nomoto}, \&
  {Umeda}}]{Hachisu1999}
{Hachisu}, I., {Kato}, M., {Nomoto}, K., \& {Umeda}, H. 1999, \apj, 519, 314

\bibitem[{{Hachisu} {et~al.}(2012){Hachisu}, {Kato}, {Saio}, \&
  {Nomoto}}]{Hachisu2012}
{Hachisu}, I., {Kato}, M., {Saio}, H., \& {Nomoto}, K. 2012, \apj, 744, 69

\bibitem[{{Hamuy} {et~al.}(2003){Hamuy}, {Phillips}, {Suntzeff}, {Maza},
  {Gonz{\'a}lez}, {Roth}, {Krisciunas}, {Morrell}, {Green}, {Persson}, \&
  {McCarthy}}]{Hamuy2003}
{Hamuy}, M., {et~al.} 2003, \nat, 424, 651

\bibitem[{{Hicken} {et~al.}(2007){Hicken}, {Garnavich}, {Prieto}, {Blondin},
  {DePoy}, {Kirshner}, \& {Parrent}}]{Hicken2007}
{Hicken}, M., {Garnavich}, P.~M., {Prieto}, J.~L., {Blondin}, S., {DePoy},
  D.~L., {Kirshner}, R.~P., \& {Parrent}, J. 2007, \apjl, 669, L17

\bibitem[{{Hillebrandt} {et~al.}(2007){Hillebrandt}, {Sim}, \&
  {R{\"o}pke}}]{Hillebrandt2007}
{Hillebrandt}, W., {Sim}, S.~A., \& {R{\"o}pke}, F.~K. 2007, \aap, 465, L17

\bibitem[{{Howell} {et~al.}(2006){Howell}, {Sullivan}, {Nugent}, {Ellis},
  {Conley}, {Le Borgne}, {Carlberg}, {Guy}, {Balam}, {Basa}, {Fouchez}, {Hook},
  {Hsiao}, {Neill}, {Pain}, {Perrett}, \& {Pritchet}}]{Howell2006}
{Howell}, D.~A., {et~al.} 2006, \nat, 443, 308

\bibitem[{{Hsiao} {et~al.}(2007){Hsiao}, {Conley}, {Howell}, {Sullivan},
  {Pritchet}, {Carlberg}, {Nugent}, \& {Phillips}}]{Hsiao2007}
{Hsiao}, E.~Y., {Conley}, A., {Howell}, D.~A., {Sullivan}, M., {Pritchet},
  C.~J., {Carlberg}, R.~G., {Nugent}, P.~E., \& {Phillips}, M.~M. 2007, \apj,
  663, 1187

\bibitem[{{Iben} \& {Tutukov}(1984)}]{Iben1984}
{Iben}, Jr., I., \& {Tutukov}, A.~V. 1984, \apjs, 54, 335

\bibitem[{{Ishiguro} {et~al.}(2015){Ishiguro}, {Kuroda}, {Hanayama},
  {Takahashi}, {Hasegawa}, {Sarugaku}, {Watanabe}, {Imai}, {Goda}, {Akitaya},
  {Takagi}, {Morihana}, {Honda}, {Arai}, {Sekiguchi}, {Oasa}, {Saito},
  {Morokuma}, {Murata}, {Nogami}, {Nagayama}, {Yanagisawa}, {Yoshida}, {Ohta},
  {Kawai}, {Miyaji}, {Fukushima}, {Watanabe}, {Opitom}, {Jehin}, {Gillon}, \&
  {Vaubaillon}}]{Ishiguro2015}
{Ishiguro}, M., {et~al.} 2015, \apjl, 798, L34

\bibitem[{{Isogai} {et~al.}(2015){Isogai}, {Arai}, {Yonehara}, {Kawakita},
  {Uemura}, \& {Nogami}}]{Isogai2015}
{Isogai}, M., {Arai}, A., {Yonehara}, A., {Kawakita}, H., {Uemura}, M., \&
  {Nogami}, D. 2015, \pasj, 67, 7

\bibitem[{{Itoh} {et~al.}(2013){Itoh}, {Fukazawa}, {Tanaka}, {Abe}, {Akitaya},
  {Arai}, {Hayashi}, {Hori}, {Isogai}, {Izumiura}, {Kawabata}, {Kawai},
  {Kuroda}, {Miyanoshita}, {Moritani}, {Morokuma}, {Nagayama}, {Nakamoto},
  {Nakata}, {Oasa}, {Ohshima}, {Ohsugi}, {Okumura}, {Saito}, {Saito}, {Sasada},
  {Sekiguchi}, {Takagi}, {Takahashi}, {Takahashi}, {Takaki}, {Uemura}, {Ueno},
  {Urakawa}, {Watanabe}, {Yamanaka}, {Yonekura}, \& {Yoshida}}]{Itoh2013}
{Itoh}, R., {et~al.} 2013, \apjl, 768, L24

\bibitem[{{Itoh} {et~al.}(2014){Itoh}, {Tanaka}, {Akitaya}, {Uemura},
  {Fukazawa}, {Inoue}, {Doi}, {Arai}, {Hanayama}, {Hashimoto}, {Hayashi},
  {Izumiura}, {Kanda}, {Kawabata}, {Kawaguchi}, {Kawai}, {Kinugasa}, {Kuroda},
  {Miyaji}, {Moritani}, {Morokuma}, {Murata}, {Nagayama}, {Oasa}, {Ohshima},
  {Ohsugi}, {Saito}, {Sakata}, {Sasada}, {Sekiguchi}, {Takagi}, {Takahashi},
  {Takaki}, {Ui}, {Watanabe}, {Yamanaka}, {Yamashita}, \& {Yoshida}}]{Itoh2014}
---. 2014, \pasj, 66, 108

\bibitem[{{Jack} {et~al.}(2012){Jack}, {Hauschildt}, \& {Baron}}]{Jack2012}
{Jack}, D., {Hauschildt}, P.~H., \& {Baron}, E. 2012, \aap, 538, A132

\bibitem[{{Kamiya} {et~al.}(2012){Kamiya}, {Tanaka}, {Nomoto}, {Blinnikov},
  {Sorokina}, \& {Suzuki}}]{Kamiya2012}
{Kamiya}, Y., {Tanaka}, M., {Nomoto}, K., {Blinnikov}, S.~I., {Sorokina},
  E.~I., \& {Suzuki}, T. 2012, \apj, 756, 191

\bibitem[{{Kawabata} {et~al.}(2000){Kawabata}, {Hirata}, {Ikeda}, {Akitaya},
  {Seki}, {Matsumura}, \& {Okazaki}}]{Kawabata2000}
{Kawabata}, K.~S., {Hirata}, R., {Ikeda}, Y., {Akitaya}, H., {Seki}, M.,
  {Matsumura}, M., \& {Okazaki}, A. 2000, \apj, 540, 429

\bibitem[{{Kawabata} {et~al.}(2008){Kawabata}, {Nagae}, {Chiyonobu}, {Tanaka},
  {Nakaya}, {Suzuki}, {Kamata}, {Miyazaki}, {Hiragi}, {Miyamoto}, {Yamanaka},
  {Arai}, {Yamashita}, {Uemura}, {Ohsugi}, {Isogai}, {Ishitobi}, \&
  {Sato}}]{Kawabata2008}
{Kawabata}, K.~S., {et~al.} 2008, in Society of Photo-Optical Instrumentation
  Engineers (SPIE) Conference Series, Vol. 7014, Society of Photo-Optical
  Instrumentation Engineers (SPIE) Conference Series

\bibitem[{{Kelly} {et~al.}(2014){Kelly}, {Fox}, {Filippenko}, {Cenko}, {Prato},
  {Schaefer}, {Shen}, {Zheng}, {Graham}, \& {Tucker}}]{Kelly2014}
{Kelly}, P.~L., {et~al.} 2014, \apj, 790, 3

\bibitem[{{Kotani} {et~al.}(2005){Kotani}, {Kawai}, {Yanagisawa}, {Watanabe},
  {Arimoto}, {Fukushima}, {Hattori}, {Inata}, {Izumiura}, {Kataoka}, {Koyano},
  {Kubota}, {Kuroda}, {Mori}, {Nagayama}, {Ohta}, {Okada}, {Okita}, {Sato},
  {Serino}, {Shimizu}, {Shimokawabe}, {Suzuki}, {Toda}, {Ushiyama}, {Yatsu},
  {Yoshida}, \& {Yoshida}}]{Kotani2005}
{Kotani}, T., {et~al.} 2005, Nuovo Cimento C Geophysics Space Physics C, 28,
  755

\bibitem[{{Krisciunas} {et~al.}(2004){Krisciunas}, {Suntzeff}, {Phillips},
  {Candia}, {Prieto}, {Antezana}, {Chassagne}, {Chen}, {Dickinson},
  {Eisenhardt}, {Espinoza}, {Garnavich}, {Gonz{\'a}lez}, {Harrison}, {Hamuy},
  {Ivanov}, {Krzemi{\'n}ski}, {Kulesa}, {McCarthy}, {Moro-Mart{\'{\i}}n},
  {Muena}, {Noriega-Crespo}, {Persson}, {Pinto}, {Roth}, {Rubenstein},
  {Stanford}, {Stringfellow}, {Zapata}, {Porter}, \&
  {Wischnjewsky}}]{Krisciunas2004}
{Krisciunas}, K., {et~al.} 2004, \aj, 128, 3034

\bibitem[{{Kuroda} {et~al.}(2015){Kuroda}, {Ishiguro}, {Watanabe}, {Akitaya},
  {Takahashi}, {Hasegawa}, {Ui}, {Kanda}, {Takaki}, {Itoh}, {Moritani}, {Imai},
  {Goda}, {Takagi}, {Morihana}, {Honda}, {Arai}, {Hanayama}, {Nagayama},
  {Nogami}, {Sarugaku}, {Murata}, {Morokuma}, {Saito}, {Oasa}, {Sekiguchi}, \&
  {Watanabe}}]{Kuroda2015}
{Kuroda}, D., {et~al.} 2015, \apj, 814, 156

\bibitem[{{Landolt}(1992)}]{Landolt1992}
{Landolt}, A.~U. 1992, \aj, 104, 340

\bibitem[{{Levanon} {et~al.}(2015){Levanon}, {Soker}, \&
  {Garc{\'{\i}}a-Berro}}]{Levanon2015}
{Levanon}, N., {Soker}, N., \& {Garc{\'{\i}}a-Berro}, E. 2015, \mnras, 447,
  2803

\bibitem[{{Li} {et~al.}(2011){Li}, {Bloom}, {Podsiadlowski}, {Miller}, {Cenko},
  {Jha}, {Sullivan}, {Howell}, {Nugent}, {Butler}, {Ofek}, {Kasliwal},
  {Richards}, {Stockton}, {Shih}, {Bildsten}, {Shara}, {Bibby}, {Filippenko},
  {Ganeshalingam}, {Silverman}, {Kulkarni}, {Law}, {Poznanski}, {Quimby},
  {McCully}, {Patel}, {Maguire}, \& {Shen}}]{Li2011}
{Li}, W., {et~al.} 2011, \nat, 480, 348

\bibitem[{{Maeda} \& {Iwamoto}(2009)}]{Maeda2009b}
{Maeda}, K., \& {Iwamoto}, K. 2009, \mnras, 394, 239

\bibitem[{{Maeda} {et~al.}(2009){Maeda}, {Kawabata}, {Li}, {Tanaka}, {Mazzali},
  {Hattori}, {Nomoto}, \& {Filippenko}}]{Maeda2009a}
{Maeda}, K., {Kawabata}, K., {Li}, W., {Tanaka}, M., {Mazzali}, P.~A.,
  {Hattori}, T., {Nomoto}, K., \& {Filippenko}, A.~V. 2009, \apj, 690, 1745

\bibitem[{{Maeda} {et~al.}(2015){Maeda}, {Nozawa}, {Nagao}, \&
  {Motohara}}]{Maeda2015}
{Maeda}, K., {Nozawa}, T., {Nagao}, T., \& {Motohara}, K. 2015, \mnras, 452,
  3281

\bibitem[{{Maeda} {et~al.}(2013){Maeda}, {Nozawa}, {Sahu}, {Minowa},
  {Motohara}, {Ueno}, {Folatelli}, {Pyo}, {Kitagawa}, {Kawabata}, {Anupama},
  {Kozasa}, {Moriya}, {Yamanaka}, {Nomoto}, {Bersten}, {Quimby}, \&
  {Iye}}]{Maeda2013}
{Maeda}, K., {et~al.} 2013, \apj, 776, 5

\bibitem[{{Marion} {et~al.}(2016){Marion}, {Brown}, {Vink{\'o}}, {Silverman},
  {Sand}, {Challis}, {Kirshner}, {Wheeler}, {Berlind}, {Brown}, {Calkins},
  {Camacho}, {Dhungana}, {Foley}, {Friedman}, {Graham}, {Howell}, {Hsiao},
  {Irwin}, {Jha}, {Kehoe}, {Macri}, {Maeda}, {Mandel}, {McCully}, {Pandya},
  {Rines}, {Wilhelmy}, \& {Zheng}}]{Marion2016}
{Marion}, G.~H., {et~al.} 2016, \apj, 820, 92

\bibitem[{{Mattila} {et~al.}(2008){Mattila}, {Meikle}, {Lundqvist},
  {Pastorello}, {Kotak}, {Eldridge}, {Smartt}, {Adamson}, {Gerardy}, {Rizzi},
  {Stephens}, \& {van Dyk}}]{Mattila2008}
{Mattila}, S., {et~al.} 2008, \mnras, 389, 141

\bibitem[{{Nagayama} {et~al.}(2003){Nagayama}, {Nagashima}, {Nakajima},
  {Nagata}, {Sato}, {Nakaya}, {Yamamuro}, {Sugitani}, \&
  {Tamura}}]{Nagayama2003}
{Nagayama}, T., {et~al.} 2003, in Society of Photo-Optical Instrumentation
  Engineers (SPIE) Conference Series, Vol. 4841, Instrument Design and
  Performance for Optical/Infrared Ground-based Telescopes, ed. M.~{Iye} \&
  A.~F.~M. {Moorwood}, 459--464

\bibitem[{{Nomoto}(1982)}]{Nomoto1982}
{Nomoto}, K. 1982, \apj, 253, 798

\bibitem[{{Nomoto} {et~al.}(1984){Nomoto}, {Thielemann}, \&
  {Yokoi}}]{Nomoto1984}
{Nomoto}, K., {Thielemann}, F., \& {Yokoi}, K. 1984, \apj, 286, 644

\bibitem[{{Nozawa} {et~al.}(2003){Nozawa}, {Kozasa}, {Umeda}, {Maeda}, \&
  {Nomoto}}]{Nozawa2003}
{Nozawa}, T., {Kozasa}, T., {Umeda}, H., {Maeda}, K., \& {Nomoto}, K. 2003,
  \apj, 598, 785

\bibitem[{{Nozawa} {et~al.}(2011){Nozawa}, {Maeda}, {Kozasa}, {Tanaka},
  {Nomoto}, \& {Umeda}}]{Nozawa2011}
{Nozawa}, T., {Maeda}, K., {Kozasa}, T., {Tanaka}, M., {Nomoto}, K., \&
  {Umeda}, H. 2011, \apj, 736, 45

\bibitem[{{Nugent} {et~al.}(2011){Nugent}, {Sullivan}, {Cenko}, {Thomas},
  {Kasen}, {Howell}, {Bersier}, {Bloom}, {Kulkarni}, {Kandrashoff},
  {Filippenko}, {Silverman}, {Marcy}, {Howard}, {Isaacson}, {Maguire},
  {Suzuki}, {Tarlton}, {Pan}, {Bildsten}, {Fulton}, {Parrent}, {Sand},
  {Podsiadlowski}, {Bianco}, {Dilday}, {Graham}, {Lyman}, {James}, {Kasliwal},
  {Law}, {Quimby}, {Hook}, {Walker}, {Mazzali}, {Pian}, {Ofek}, {Gal-Yam}, \&
  {Poznanski}}]{Nugent2011}
{Nugent}, P.~E., {et~al.} 2011, \nat, 480, 344

\bibitem[{{Ohtani} {et~al.}(1998){Ohtani}, {Ishigaki}, {Maemura}, {Hayashi},
  {Sasaki}, {Ozaki}, {Hattori}, {Aoki}, \& {Sugai}}]{Ohtani1998}
{Ohtani}, H., {et~al.} 1998, in Society of Photo-Optical Instrumentation
  Engineers (SPIE) Conference Series, Vol. 3355, Optical Astronomical
  Instrumentation, ed. S.~{D'Odorico}, 750--761

\bibitem[{{Pakmor} {et~al.}(2010){Pakmor}, {Kromer}, {R{\"o}pke}, {Sim},
  {Ruiter}, \& {Hillebrandt}}]{Pakmor2010}
{Pakmor}, R., {Kromer}, M., {R{\"o}pke}, F.~K., {Sim}, S.~A., {Ruiter}, A.~J.,
  \& {Hillebrandt}, W. 2010, \nat, 463, 61

\bibitem[{{Pakmor} {et~al.}(2012){Pakmor}, {Kromer}, {Taubenberger}, {Sim},
  {R{\"o}pke}, \& {Hillebrandt}}]{Pakmor2012}
{Pakmor}, R., {Kromer}, M., {Taubenberger}, S., {Sim}, S.~A., {R{\"o}pke},
  F.~K., \& {Hillebrandt}, W. 2012, \apjl, 747, L10

\bibitem[{{Parrent} \& {Howell}(2012)}]{Parrent2012}
{Parrent}, J.~T., \& {Howell}, D.~A. 2012, Central Bureau Electronic Telegrams,
  3174, 2

\bibitem[{{Parrent} {et~al.}(2016){Parrent}, {Howell}, {Fesen}, {Parker},
  {Bianco}, {Dilday}, {Sand}, {Valenti}, {Vink{\'o}}, {Berlind}, {Challis},
  {Milisavljevic}, {Sanders}, {Marion}, {Wheeler}, {Brown}, {Calkins},
  {Friesen}, {Kirshner}, {Pritchard}, {Quimby}, \& {Roming}}]{Parrent2016}
{Parrent}, J.~T., {et~al.} 2016, \mnras

\bibitem[{{Patat}(2005)}]{Patat2005}
{Patat}, F. 2005, \mnras, 357, 1161

\bibitem[{{Patat} {et~al.}(2006){Patat}, {Benetti}, {Cappellaro}, \&
  {Turatto}}]{Patat2006}
{Patat}, F., {Benetti}, S., {Cappellaro}, E., \& {Turatto}, M. 2006, \mnras,
  369, 1949

\bibitem[{{Patat} {et~al.}(2007){Patat}, {Chandra}, {Chevalier}, {Justham},
  {Podsiadlowski}, {Wolf}, {Gal-Yam}, {Pasquini}, {Crawford}, {Mazzali},
  {Pauldrach}, {Nomoto}, {Benetti}, {Cappellaro}, {Elias-Rosa}, {Hillebrandt},
  {Leonard}, {Pastorello}, {Renzini}, {Sabbadin}, {Simon}, \&
  {Turatto}}]{Patat2007}
{Patat}, F., {et~al.} 2007, Science, 317, 924

\bibitem[{{Perlmutter} {et~al.}(1999){Perlmutter}, {Aldering}, {Goldhaber},
  {Knop}, {Nugent}, {Castro}, {Deustua}, {Fabbro}, {Goobar}, {Groom}, {Hook},
  {Kim}, {Kim}, {Lee}, {Nunes}, {Pain}, {Pennypacker}, {Quimby}, {Lidman},
  {Ellis}, {Irwin}, {McMahon}, {Ruiz-Lapuente}, {Walton}, {Schaefer}, {Boyle},
  {Filippenko}, {Matheson}, {Fruchter}, {Panagia}, {Newberg}, {Couch}, \& {The
  Supernova Cosmology Project}}]{Perlmutter1999}
{Perlmutter}, S., {et~al.} 1999, \apj, 517, 565

\bibitem[{{Persson} {et~al.}(1998){Persson}, {Murphy}, {Krzeminski}, {Roth}, \&
  {Rieke}}]{Persson1998}
{Persson}, S.~E., {Murphy}, D.~C., {Krzeminski}, W., {Roth}, M., \& {Rieke},
  M.~J. 1998, \aj, 116, 2475

\bibitem[{{Pfannes} {et~al.}(2010){Pfannes}, {Niemeyer}, \&
  {Schmidt}}]{Pfannes2010}
{Pfannes}, J.~M.~M., {Niemeyer}, J.~C., \& {Schmidt}, W. 2010, \aap, 509, A75

\bibitem[{{Phillips}(1993)}]{Phillips1993}
{Phillips}, M.~M. 1993, \apjl, 413, L105

\bibitem[{{Phillips} {et~al.}(1999){Phillips}, {Lira}, {Suntzeff}, {Schommer},
  {Hamuy}, \& {Maza}}]{Phillips1999}
{Phillips}, M.~M., {Lira}, P., {Suntzeff}, N.~B., {Schommer}, R.~A., {Hamuy},
  M., \& {Maza}, J. 1999, \aj, 118, 1766

\bibitem[{{Pozzo} {et~al.}(2004){Pozzo}, {Meikle}, {Fassia}, {Geballe},
  {Lundqvist}, {Chugai}, \& {Sollerman}}]{Pozzo2004}
{Pozzo}, M., {Meikle}, W.~P.~S., {Fassia}, A., {Geballe}, T., {Lundqvist}, P.,
  {Chugai}, N.~N., \& {Sollerman}, J. 2004, \mnras, 352, 457

\bibitem[{{Prieto} {et~al.}(2006){Prieto}, {Rest}, \& {Suntzeff}}]{Prieto2006}
{Prieto}, J.~L., {Rest}, A., \& {Suntzeff}, N.~B. 2006, \apj, 647, 501

\bibitem[{{Quinn} {et~al.}(2006){Quinn}, {Garnavich}, {Li}, {Panagia}, {Riess},
  {Schmidt}, \& {Della Valle}}]{Quinn2006}
{Quinn}, J.~L., {Garnavich}, P.~M., {Li}, W., {Panagia}, N., {Riess}, A.,
  {Schmidt}, B.~P., \& {Della Valle}, M. 2006, \apj, 652, 512

\bibitem[{{Riess} {et~al.}(1998){Riess}, {Filippenko}, {Challis},
  {Clocchiatti}, {Diercks}, {Garnavich}, {Gilliland}, {Hogan}, {Jha},
  {Kirshner}, {Leibundgut}, {Phillips}, {Reiss}, {Schmidt}, {Schommer},
  {Smith}, {Spyromilio}, {Stubbs}, {Suntzeff}, \& {Tonry}}]{Riess1998}
{Riess}, A.~G., {et~al.} 1998, \aj, 116, 1009

\bibitem[{{Sakon} {et~al.}(2009){Sakon}, {Onaka}, {Wada}, {Ohyama}, {Kaneda},
  {Ishihara}, {Tanab{\'e}}, {Minezaki}, {Yoshii}, {Tominaga}, {Nomoto},
  {Nozawa}, {Kozasa}, {Tanaka}, {Suzuki}, {Umeda}, {Ohyabu}, {Usui},
  {Matsuhara}, {Nakagawa}, \& {Murakami}}]{Sakon2009}
{Sakon}, I., {et~al.} 2009, \apj, 692, 546

\bibitem[{{Scalzo} {et~al.}(2012){Scalzo}, {Aldering}, {Antilogus}, {Aragon},
  {Bailey}, {Baltay}, {Bongard}, {Buton}, {Canto}, {Cellier-Holzem},
  {Childress}, {Chotard}, {Copin}, {Fakhouri}, {Gangler}, {Guy}, {Hsiao},
  {Kerschhaggl}, {Kowalski}, {Nugent}, {Paech}, {Pain}, {Pecontal}, {Pereira},
  {Perlmutter}, {Rabinowitz}, {Rigault}, {Runge}, {Smadja}, {Tao}, {Thomas},
  {Weaver}, {Wu}, \& {Nearby Supernova Factory}}]{Scalzo2012}
{Scalzo}, R., {et~al.} 2012, \apj, 757, 12

\bibitem[{{Scalzo} {et~al.}(2010){Scalzo}, {Aldering}, {Antilogus}, {Aragon},
  {Bailey}, {Baltay}, {Bongard}, {Buton}, {Childress}, {Chotard}, {Copin},
  {Fakhouri}, {Gal-Yam}, {Gangler}, {Hoyer}, {Kasliwal}, {Loken}, {Nugent},
  {Pain}, {P{\'e}contal}, {Pereira}, {Perlmutter}, {Rabinowitz}, {Rau},
  {Rigaudier}, {Runge}, {Smadja}, {Tao}, {Thomas}, {Weaver}, \&
  {Wu}}]{Scalzo2010}
{Scalzo}, R.~A., {et~al.} 2010, \apj, 713, 1073

\bibitem[{{Schaefer}(1987)}]{Schaefer1987}
{Schaefer}, B.~E. 1987, \apjl, 323, L47

\bibitem[{{Schlafly} \& {Finkbeiner}(2011)}]{Schlafly2011}
{Schlafly}, E.~F., \& {Finkbeiner}, D.~P. 2011, \apj, 737, 103

\bibitem[{{Silverman} {et~al.}(2011){Silverman}, {Ganeshalingam}, {Li},
  {Filippenko}, {Miller}, \& {Poznanski}}]{Silverman2011}
{Silverman}, J.~M., {Ganeshalingam}, M., {Li}, W., {Filippenko}, A.~V.,
  {Miller}, A.~A., \& {Poznanski}, D. 2011, \mnras, 410, 585

\bibitem[{{Simon} {et~al.}(2009){Simon}, {Gal-Yam}, {Gnat}, {Quimby},
  {Ganeshalingam}, {Silverman}, {Blondin}, {Li}, {Filippenko}, {Wheeler},
  {Kirshner}, {Patat}, {Nugent}, {Foley}, {Vogt}, {Butler}, {Peek},
  {Rosolowsky}, {Herczeg}, {Sauer}, \& {Mazzali}}]{Simon2009}
{Simon}, J.~D., {et~al.} 2009, \apj, 702, 1157

\bibitem[{{Sollerman} {et~al.}(2004){Sollerman}, {Lindahl}, {Kozma}, {Challis},
  {Filippenko}, {Fransson}, {Garnavich}, {Leibundgut}, {Li}, {Lundqvist},
  {Milne}, {Spyromilio}, \& {Kirshner}}]{Sollerman2004}
{Sollerman}, J., {et~al.} 2004, \aap, 428, 555

\bibitem[{{Sparks} {et~al.}(1999){Sparks}, {Macchetto}, {Panagia}, {Boffi},
  {Branch}, {Hazen}, \& {Della Valle}}]{Sparks1999}
{Sparks}, W.~B., {Macchetto}, F., {Panagia}, N., {Boffi}, F.~R., {Branch}, D.,
  {Hazen}, M.~L., \& {Della Valle}, M. 1999, \apj, 523, 585

\bibitem[{{Stanishev} {et~al.}(2007){Stanishev}, {Goobar}, {Benetti}, {Kotak},
  {Pignata}, {Navasardyan}, {Mazzali}, {Amanullah}, {Garavini}, {Nobili},
  {Qiu}, {Elias-Rosa}, {Ruiz-Lapuente}, {Mendez}, {Meikle}, {Patat},
  {Pastorello}, {Altavilla}, {Gustafsson}, {Harutyunyan}, {Iijima},
  {Jakobsson}, {Kichizhieva}, {Lundqvist}, {Mattila}, {Melinder}, {Pavlenko},
  {Pavlyuk}, {Sollerman}, {Tsvetkov}, {Turatto}, \&
  {Hillebrandt}}]{Stanishev2007}
{Stanishev}, V., {et~al.} 2007, \aap, 469, 645

\bibitem[{{Stetson}(1987)}]{Stetson1987}
{Stetson}, P.~B. 1987, \pasp, 99, 191

\bibitem[{{Stritzinger} {et~al.}(2006){Stritzinger}, {Mazzali}, {Sollerman}, \&
  {Benetti}}]{Stritzinger2006}
{Stritzinger}, M., {Mazzali}, P.~A., {Sollerman}, J., \& {Benetti}, S. 2006,
  \aap, 460, 793

\bibitem[{{Tanaka} {et~al.}(2011){Tanaka}, {Mazzali}, {Stanishev}, {Maurer},
  {Kerzendorf}, \& {Nomoto}}]{Tanaka2011}
{Tanaka}, M., {Mazzali}, P.~A., {Stanishev}, V., {Maurer}, I., {Kerzendorf},
  W.~E., \& {Nomoto}, K. 2011, \mnras, 410, 1725

\bibitem[{{Tanaka} {et~al.}(2010){Tanaka}, {Kawabata}, {Yamanaka}, {Maeda},
  {Hattori}, {Aoki}, {Nomoto}, {Iye}, {Sasaki}, {Mazzali}, \&
  {Pian}}]{Tanaka2010a}
{Tanaka}, M., {et~al.} 2010, \apj, 714, 1209

\bibitem[{{Tanikawa} {et~al.}(2015){Tanikawa}, {Nakasato}, {Sato}, {Nomoto},
  {Maeda}, \& {Hachisu}}]{Tanikawa2015}
{Tanikawa}, A., {Nakasato}, N., {Sato}, Y., {Nomoto}, K., {Maeda}, K., \&
  {Hachisu}, I. 2015, \apj, 807, 40

\bibitem[{{Taubenberger} {et~al.}(2011){Taubenberger}, {Benetti}, {Childress},
  {Pakmor}, {Hachinger}, {Mazzali}, {Stanishev}, {Elias-Rosa}, {Agnoletto},
  {Bufano}, {Ergon}, {Harutyunyan}, {Inserra}, {Kankare}, {Kromer},
  {Navasardyan}, {Nicolas}, {Pastorello}, {Prosperi}, {Salgado}, {Sollerman},
  {Stritzinger}, {Turatto}, {Valenti}, \& {Hillebrandt}}]{Taubenberger2011}
{Taubenberger}, S., {et~al.} 2011, \mnras, 61

\bibitem[{{Theureau} {et~al.}(1998){Theureau}, {Bottinelli}, {Coudreau-Durand},
  {Gouguenheim}, {Hallet}, {Loulergue}, {Paturel}, \&
  {Teerikorpi}}]{Theureau1998}
{Theureau}, G., {Bottinelli}, L., {Coudreau-Durand}, N., {Gouguenheim}, L.,
  {Hallet}, N., {Loulergue}, M., {Paturel}, G., \& {Teerikorpi}, P. 1998,
  \aaps, 130, 333

\bibitem[{{Wang}(2005)}]{LWang2005}
{Wang}, L. 2005, \apjl, 635, L33

\bibitem[{{Wang} {et~al.}(2008){Wang}, {Li}, {Filippenko}, {Foley}, {Smith}, \&
  {Wang}}]{XWang2008b}
{Wang}, X., {Li}, W., {Filippenko}, A.~V., {Foley}, R.~J., {Smith}, N., \&
  {Wang}, L. 2008, \apj, 677, 1060

\bibitem[{{Wang} {et~al.}(2006){Wang}, {Wang}, {Pain}, {Zhou}, \&
  {Li}}]{XWang2006}
{Wang}, X., {Wang}, L., {Pain}, R., {Zhou}, X., \& {Li}, Z. 2006, \apj, 645,
  488

\bibitem[{{Wang} {et~al.}(2009){Wang}, {Li}, {Filippenko}, {Foley}, {Kirshner},
  {Modjaz}, {Bloom}, {Brown}, {Carter}, {Friedman}, {Gal-Yam}, {Ganeshalingam},
  {Hicken}, {Krisciunas}, {Milne}, {Silverman}, {Suntzeff}, {Wood-Vasey},
  {Cenko}, {Challis}, {Fox}, {Kirkman}, {Li}, {Li}, {Malkan}, {Moore},
  {Reitzel}, {Rich}, {Serduke}, {Shang}, {Steele}, {Swift}, {Tao}, {Wong}, \&
  {Zhang}}]{XWang2009a}
{Wang}, X., {et~al.} 2009, \apj, 697, 380

\bibitem[{{Watanabe} {et~al.}(2012){Watanabe}, {Takahashi}, {Sato}, {Watanabe},
  {Fukuhara}, {Hamamoto}, \& {Ozaki}}]{Watanabe2012}
{Watanabe}, M., {Takahashi}, Y., {Sato}, M., {Watanabe}, S., {Fukuhara}, T.,
  {Hamamoto}, K., \& {Ozaki}, A. 2012, in Society of Photo-Optical
  Instrumentation Engineers (SPIE) Conference Series, Vol. 8446, Society of
  Photo-Optical Instrumentation Engineers (SPIE) Conference Series

\bibitem[{{Webbink}(1984)}]{Webbink1984}
{Webbink}, R.~F. 1984, \apj, 277, 355

\bibitem[{{Woudt} {et~al.}(2009){Woudt}, {Steeghs}, {Karovska}, {Warner},
  {Groot}, {Nelemans}, {Roelofs}, {Marsh}, {Nagayama}, {Smits}, \&
  {O'Brien}}]{Woudt2009}
{Woudt}, P.~A., {et~al.} 2009, \apj, 706, 738

\bibitem[{{Yamanaka} {et~al.}(2009){Yamanaka}, {Kawabata}, {Kinugasa},
  {Tanaka}, {Imada}, {Maeda}, {Nomoto}, {Arai}, {Chiyonobu}, {Fukazawa},
  {Hashimoto}, {Honda}, {Ikejiri}, {Itoh}, {Kamata}, {Kawai}, {Komatsu},
  {Konishi}, {Kuroda}, {Miyamoto}, {Miyazaki}, {Nagae}, {Nakaya}, {Ohsugi},
  {Omodaka}, {Sakai}, {Sasada}, {Suzuki}, {Taguchi}, {Takahashi}, {Tanaka},
  {Uemura}, {Yamashita}, {Yanagisawa}, \& {Yoshida}}]{Yamanaka2009a}
{Yamanaka}, M., {et~al.} 2009, \apjl, 707, L118

\bibitem[{{Yamanaka} {et~al.}(2014){Yamanaka}, {Maeda}, {Kawabata}, {Tanaka},
  {Takaki}, {Ueno}, {Masumoto}, {Kawabata}, {Itoh}, {Moritani}, {Akitaya},
  {Arai}, {Honda}, {Nishiyama}, {Kabashima}, {Matsumoto}, {Nogami}, \&
  {Yoshida}}]{Yamanaka2014}
---. 2014, \apjl, 782, L35

\bibitem[{{Yamanaka} {et~al.}(2015){Yamanaka}, {Maeda}, {Kawabata}, {Tanaka},
  {Tominaga}, {Akitaya}, {Nagayama}, {Kuroda}, {Takahashi}, {Saito},
  {Yanagisawa}, {Fukui}, {Miyanoshita}, {Watanabe}, {Arai}, {Isogai},
  {Hattori}, {Hanayama}, {Itoh}, {Ui}, {Takaki}, {Ueno}, {Yoshida}, {Ali},
  {Essam}, {Ozaki}, {Nakao}, {Hamamoto}, {Nogami}, {Morokuma}, {Oasa},
  {Izumiura}, \& {Sekiguchi}}]{Yamanaka2015}
---. 2015, \apj, 806, 191

\bibitem[{{Yanagisawa} {et~al.}(2010){Yanagisawa}, {Kuroda}, {Yoshida},
  {Shimizu}, {Nagayama}, {Toda}, {Ohta}, \& {Kawai}}]{Yanagisawa2010}
{Yanagisawa}, K., {Kuroda}, D., {Yoshida}, M., {Shimizu}, Y., {Nagayama}, S.,
  {Toda}, H., {Ohta}, K., \& {Kawai}, N. 2010, in American Institute of Physics
  Conference Series, Vol. 1279, American Institute of Physics Conference
  Series, ed. N.~{Kawai} \& S.~{Nagataki}, 466--468

\bibitem[{{Yatsu} {et~al.}(2015){Yatsu}, {Kataoka}, {Takahashi}, {Tachibana},
  {Kawai}, {Shibata}, {Pike}, {Yoshii}, {Arimoto}, {Saito}, {Nakamori},
  {Sekiguchi}, {Kuroda}, {Yanagisawa}, {Hanayama}, {Watanabe}, {Hamamoto},
  {Nakao}, {Ozaki}, {Motohara}, {Konishi}, {Tateuchi}, {Matsunaga}, {Morokuma},
  {Nagayama}, {Murata}, {Akitaya}, {Yoshida}, {Ali}, {Essam Mohamed}, {Isogai},
  {Arai}, {Takahashi}, {Hashimoto}, {Miyanoshita}, {Omodaka}, {Takahashi},
  {Tokimasa}, {Matsuda}, {Okumura}, {Nishiyama}, {Urakawa}, {Nogami}, {Oasa},
  \& {OISTER Team}}]{Yatsu2015}
{Yatsu}, Y., {et~al.} 2015, \apj, 802, 84

\bibitem[{{Yoon} {et~al.}(2007){Yoon}, {Podsiadlowski}, \&
  {Rosswog}}]{Yoon2007}
{Yoon}, S.-C., {Podsiadlowski}, P., \& {Rosswog}, S. 2007, \mnras, 380, 933

\bibitem[{{Yoshida}(2005)}]{Yoshida2005}
{Yoshida}, M. 2005, Journal of Korean Astronomical Society, 38, 117

\bibitem[{{Yuan} {et~al.}(2010){Yuan}, {Quimby}, {Wheeler}, {Vink{\'o}},
  {Chatzopoulos}, {Akerlof}, {Kulkarni}, {Miller}, {McKay}, \&
  {Aharonian}}]{Yuan2010}
{Yuan}, F., {et~al.} 2010, \apj, 715, 1338

\end{thebibliography}
\end{document}